\documentclass[a4paper,12pt]{article}
\pdfoutput=1
\usepackage{amssymb}
\usepackage{epsfig}
\usepackage{graphicx}
\usepackage{slashed}
\usepackage{mathtools}
\usepackage{braket}
\usepackage{dsfont}

\makeatletter

\@addtoreset{equation}{section}
\makeatother
\def\be{\begin{equation}}
\def\ee{\end{equation}}
\def\bea{\begin{eqnarray}}
\def\eea{\end{eqnarray}}

\def\({\left(}
\def\){\right)}
\def\<{\left<}
\def\>{\right>}

\def\tr{{\mbox{tr}}}
\def\be{\begin{equation}}
\def\ee{\end{equation}}
\def\bea{\begin{eqnarray*}}
\def\eea{\end{eqnarray*}}
\def\ben{\begin{eqnarray}}
\def\een{\end{eqnarray}}
\def\({\left(}
\def\){\right)}
\def\<{\left<}
\def\>{\right>}
\def\!{\right|}
\def\|{\left|}

\def\[{\left[}
\def\]{\right]}

\def\+{\bar}
\def\mb{\mathbb}
\def\tr{{\mbox{tr}}}
\def\Tr{{\mbox{Tr}}}
\def\D{{\cal{D}}}
\def\L{{\cal{L}}}
\def\t{\widetilde}
\def\A{{\cal{A}}}

\def\N{{\cal{N}}}

\def\F{{\cal{F}}}
\def\G{{\cal{G}}}

\def\H{{\cal{H}}}

\def\L{{\cal{L}}}

\def\F{{\cal{F}}}

\def\1{\mathds{1}}

\def\l{{{\ell}}}

\def\h{\widehat}

\def\b{\overline}

\def\Vol{{\mbox{Vol}}}
\def\Ker{{\mbox{Ker}}}

\def\lk{{\ell{\mbox{k}}}}

\begin{document}

\setlength{\unitlength}{1mm}

\pagestyle{empty}
\vskip-10pt
\vskip-10pt
\hfill 
\begin{center}
\vskip 3truecm
{\Large \bf
Five-dimensional fermionic Chern-Simons theory}
\vskip 2truecm
{\large \textsc{
Dongsu Bak$^{\tt a}$ and  Andreas Gustavsson$^{\tt b}$}}
\vspace{1cm} 
\begin{center} 
\it  a) Physics Department, University of Seoul,  Seoul 02504, Korea\\
b) Department of Physics and Astronomy, Uppsala University,\\
Box 516, SE-75120 Uppsala, Sweden
\end{center}
\vskip 0.7truecm
\begin{center}
(\tt dsbak@uos.ac.kr, agbrev@gmail.com)
\end{center}
\end{center}
\vskip 2truecm
{\abstract{We study 5d fermionic CS theory with a fermionic 2-form gauge potential. This theory can be obtained from 5d MSYM theory by performing the maximal topological twist. We put the theory on a five-manifold and compute the partition function. We find that it is a topological quantity, which involves the Ray-Singer torsion of the five-manifold. For abelian gauge group we consider the uplift to the 6d theory and find a mismatch between the 5d partition function and the 6d index, due to the nontrivial dimensional reduction of a selfdual two-form gauge field on a circle. We also discuss an application of the 5d theory to generalized knots made of 2d sheets embedded in 5d.}}

\vfill
\vskip4pt
\eject
\pagestyle{plain}

\section{Introduction}
Chern-Simons theory in 3d whose classical action is given by
\ben
\frac{k}{4\pi} \int \tr\(A\wedge dA - \frac{2i}{3} A\wedge A\wedge A\)\label{3dCS}
\een
has a long history. The seminal paper \cite{Witten:1988hf} obtained the exact result for the partition function for $S^3$ by indirect methods. Later exact results have been obtained in \cite{Beasley:2005vf, Blau:2006gh, Kallen:2011ny} by various methods (nonabelian localization \cite{Witten:1992xu}, abelianization, supersymmetric localization \cite{Pestun:2007rz}) on a large class of three-manifolds. There have also been many works that have aimed to match such exact results with corresponding perturbative results in the large $k$ limit \cite{Freed:1991wd, Rozansky:1993zx, Adams:1995xj, Adams:1995np, Adams:1996hi, Adams:1997zc}. Thus CS theory enables one to test path integral methods against known exact results.

We may generalize abelian CS theory to $2p+1$ dimensions by taking the gauge potential to be a $p$-form. When $p$ is odd, the gauge field is bosonic. However, when $p$ is even, a bosonic gauge field leads to a CS term that is a total derivative since $A_p \wedge dA_p = \frac{1}{2}d(A_p \wedge A_p)$. For even $p$ we shall therefore take the $p$-form gauge field to be fermionic and then we have a fermionic CS theory or FCS theory for short. In the first few dimensions these CS and FCS actions, in Lorentzian signature and with canonical normalizations, are given by
\ben
S_{1d} &=& \frac{i}{2} \int \psi_0 \wedge d\psi_0\cr
S_{3d} &=& \frac{1}{2} \int A_1 \wedge dA_1\cr
S_{5d} &=& \frac{i}{2} \int \psi_2 \wedge d\psi_2\cr
S_{7d} &=& \frac{1}{2} \int A_3 \wedge dA_3\label{dCS}
\een
The most general form of the gauge symmetry variations are
\bea
\delta \psi_0 &=& (\chi_0)_0\cr
\delta A_1 &=& d\lambda_0 + (\lambda_1)_0\cr
\delta \psi_2 &=& d\chi_1 + (\chi_2)_0\cr
\delta A_3 &=& d\lambda_2 + (\lambda_3)_0
\eea
In addition to the usual exact forms, we shall also include the harmonic forms $(\chi_p)_0$ and $(\lambda_p)_0$ in order to have the most general closed forms by the Hodge decomposition \cite{Blau:1989bq}. The fact that abelian CS theories in various dimensions form a sequence (\ref{dCS}) suggests that they could have some common features.

It is in general a quite difficult problem to generalize abelian higher rank gauge fields to nonabelian gauge groups. However, in 5d we automatically solve this problem since 5d FCS is obtained from 5d MSYM theory by performing the maximal twist. By this twist the $SO(5)$ R-symmetry is identified with the $SO(5)$ Lorentz symmetry \cite{Bak:2015hba, Park}. The twist gives one scalar nilpotent supercharge, which we can identify as the BRST charge associated with the two-form gauge symmetry, and the action can be interpreted as a BRST gauge fixed action for nonabelian 5d FCS theory. 

For 3d CS on lens space $S^3/\mb{Z}_p = L(p;1)$, the exact partition function is known. From this exact result we can extract the perturbative expansion in $1/k$. For gauge group $G=SU(2)$, the resulting perturbative expansion for $p$ odd, is\footnote{If $p$ is even, then at $\l=p/2$ we get $U$$=$ diag$(-1,-1)$ which commutes with all group elements in $SU(2)$. In this case we should probably have the gauge group as $SU(2)/\mb{Z}_2$ and identify this holonomy with the holonomy at $\l =0$. We then do not count it since we only count gauge inequivalent holonomies. So for $p$ even, we sum over the holonomy sectors $\l =0,1,...,p/2-1$ and then this will again be in agreement with the general formula in \cite{Beasley:2005vf}.}
\bea
Z &=& e^\frac{i\pi p}{4} \pi \sqrt{2} \frac{1}{\(iKp\)^{3/2}} + e^\frac{i \pi p}{4} \frac{2\sqrt{2}}{\sqrt{iKp}} \sum_{\l=1}^{\frac{p-1}{2}} e^\frac{2\pi i K \l^2}{p} \(\sin\frac{2\pi\l}{p}\)^2
\eea
where $K = k+2$. This agrees with the perturbative expansion in \cite{Rozansky:1993zx}. We obtained this result from the exact result presented in \cite{Beasley:2005vf} by expanding it out in powers of $1/k$ but where we suppress the next to leading orders in each sector labeled by $\l=0,1,...,\frac{p-1}{2}$. This result can be rewritten in the form
\ben
Z &=& e^{\frac{i\pi}{4} \dim(G) \eta_{grav}} \[\frac{1}{\Vol\(H_{A^{(0)}}\)}  \sqrt{\tau_{0,SU(2)}} + \sum_{\ell=1}^{\frac{p-1}{2}} \frac{1}{\Vol\(H_{A^{(\ell)}}\)}  e^\frac{2\pi i K \ell^2}{p} \sqrt{\tau_{\l,SU(2)}}\] \label{3dCS}
\een
Here $H_{A^{(\l)}}$ denotes the unbroken gauge group by the gauge field background and $\tau_{\l,SU(2)}$ denotes the Ray-Singer torsion of $L(p;1)$ associated with $SU(2)$ gauge group and the holonomy labeled by $\l=0,...,p-1$. For the lens space $L(p;q_1,\cdots,q_{N-1})=S^{2N-1}/\mb{Z}_p$ we have
\bea
\tau_{0,SU(2)} &=& (\tau_0)^3\cr
\tau_{\l,SU(2)} &=& \tau_0 \tau_{2\l} \tau_{-2\l}\qquad (\l\neq 0)
\eea
where
\bea
\tau_0 &=& \frac{1}{p^{N-1}}\cr
\tau_{\l} &=& \left|2^N \sin \frac{\pi q_1^{-1} \l}{p} \cdots \sin \frac{\pi q_{N-1}^{-1} \l}{p} \sin \frac{\pi \l}{p}\right|
\eea
To get the torsion for $L(p;1)$, we shall put $N=2$ and $q=q^{-1} = 1$. The unbroken gauge groups are $H_{A^{(0)}} = SU(2)$ and $H_{A^{(\l)}} = U(1)$, whose volumes are
\bea
\Vol(U(1)) &=& 2\pi r\cr
\Vol(SU(2)) &=& 2\pi^2 r^3
\eea
We note that $U(1)$ corresponds to the equator of $SU(2) = S^3$. The radius shall be chosen as
\bea
r &=& \sqrt{\frac{i}{\pi}} \sqrt{\frac{K}{2\pi}}
\eea
in order to match with the exact result. We can see why this value of the radius is natural up to the factor $\sqrt{\frac{i}{\pi}}$ as follows. We need to rescale $A = \sqrt{\frac{K}{2\pi}} A_{can}$ to get a canonically normalized action. This can be achieved by rescaling the generators of $SU(2)$ by the factor $\sqrt{\frac{K}{2\pi}}$. The overall factor $e^{\frac{i\pi}{4} \dim(G) \eta_{grav}}$ is the remnant of the eta-invariant phase shift that results in the famous shift of the CS level from $k$ to $K = k+2$.

Next we consider 1d FCS on $L(p) = S^1/\mb{Z}_p$ with gauge group $SU(2)$ and the following action in Euclidean signature
\bea
S &=& \frac{k}{8\pi} \int_0^{2\pi} dx^0 \tr(\psi D_0 \psi)
\eea
To compute the Witten index from the path integral, we do not need to use a Faddeev-Popov gauge fixing procedure since we can directly specify the gauge inequivalent gauge field configurations, which are classified by the holonomies
\bea
P \exp i \int_0^{2\pi} dx^0 A_0 \begin{pmatrix}
1 & 0\\
0 & -1
\end{pmatrix} = \begin{pmatrix}
e^{\frac{2\pi i \l}{p}} & 0\\
0 & e^{-\frac{2\pi i \l}{p}}
\end{pmatrix}
\eea
We can pick the gauge inequivalent gauge fields as
\bea
A_0 &=& \frac{\l}{p} \begin{pmatrix}
1 & 0\\
0 & -1
\end{pmatrix} 
\eea
and the path integral reduces to a discrete sum over $\l$. This sum is presented in Eq.~(\ref{a sum}). But we can also carry out the standard Faddeev-Popov procedure and if we do that, then we are led to the result\footnote{As we will see, this result is correct only up to an overall phase factor.}
\bea
I &=& \frac{1}{\Vol(H_{A^{(0)}})} \sqrt{\tau_{0,SU(2)}} + \frac{1}{\Vol(H_{A^{(\l)}})} \sum_{\l=1}^{\frac{p-1}{2}} \sqrt{\tau_{\l,SU(2)}}
\eea
where
\ben
\tau_{0,SU(2)} &=& (\tau_0)^3\cr
\tau_{\l,SU(2)} &=& \tau_0 \tau_{2\l} \tau_{-2\l}\qquad (\l\neq 0)\label{SU2RS}
\een
with
\bea
\tau_0 &=& 1\cr
\tau_{\l} &=& \left|2 \sin \frac{\pi \l}{p}\right|
\eea
As we will show, the two expressions can be made to agree by taking the following radius for the $SU(2)$ gauge group,
\bea
r &=& \sqrt{\frac{i}{\pi}} \sqrt{\frac{k}{2\pi}}
\eea
We notice that this radius takes the same form as we saw for 3d CS.

If we use our conjectured similarity between FCS theories in various dimensions, then we are led to the partition function 
\ben
Z &=& \frac{1}{\Vol\(H_{A^{(0)}}\)} \sqrt{\tau_{0,SU(2)}} + \sum_{\ell=1}^{\frac{p-1}{2}} \frac{1}{\Vol\(H_{A^{(\ell)}}\)} \sqrt{\tau_{\l,SU(2)}} \label{5dCS}
\een
for 5d FCS with gauge group $G=SU(2)$. We conjecture that the volume factors are on the same form as for 1d FCS when the 5d FCS action is canonically normalized, but we have not been able to explicitly compute the radius $r$ for this case. Gauge fixing amounts to adding BRST exact terms that we can also obtain by twisting of 5d MSYM. We will partly be able to confirm our conjecture by a localization computation in section 4.1. 

We see that FCS theories differ from CS theories in many ways. Of course we do not know much about CS theories in other dimensions than three. For FCS, the partition function is one-loop exact. There is no phase factor multiplying the contributions from the various holonomy sectors, and there is no Chern-Simons level $k$ that can take arbitrary integer values for FCS. 

The paper is organized as follows. In section $2$ we construct explicit solutions for flat gauge fields on lens spaces in 3d and in 5d. In section $3$ we compute the Witten index for 1d FCS. In section $4$ we compute the partition function for 5d FCS. In section $5$ we discuss applications to higher dimensional knots. In sections $6$ and $7$ we obtain the mismatch between the 5d partition function and the 6d Witten index that is related to the Ray-Singer torsion. 

The appendices contain further details which makes the paper self-contained. In appendix A we review the definition and basic properties of the Ray-Singer torsion. In appendix A.1 we compute the Ray-Singer torsion on $L(p;1,1)$ in the trivial holonomy sector. In appendix B we present the Minakshisundaram-Pleijel theorem, which we use throughout the paper. In appendices C and D we address the problem of how to properly remove ghost zero modes. In appendix E we review what we need from 3d CS perturbation theory. In appendix F we present further details regarding the dimensional reduction on a circle and the mismatch related to the Ray-Singer torsion in various dimensions.

\section{Flat gauge fields on lens spaces}
For the lens space $L_{2N-1}(g) = \t L(p;q_1,\dots,q_N) = S^{2N-1}/\mb{Z}_p$, the generator $g$ of $\mb{Z}_p$ acts on $\mb{C}^N$ as
\bea
g(z_1,\dots,z_N) = (e^{2\pi i q_1/p} z_1,\dots,e^{2\pi i q_N/p}z_N)
\eea
For twisted boundary conditions
\bea
g \phi &=& \omega(g) \phi
\eea
or explicitly
\bea
\phi(e^{2\pi i q_1/p} z_1,\dots,e^{2\pi i q_N/p}z_N) &=& e^{2\pi i q_1 \ell/p} \phi(z_1,z_2,z_3)
\eea
where we characterize $\omega$ by the integer $\l$, the Ray-Singer torsion was first computed by Ray by assuming $N\geq 2$ in \cite{Ray}. A direct computation has been made in \cite{Nash:1992sf, Friedmann:2002ty}. Lecture notes on the Ray-Singer torsion are \cite{Bunke, Mnev}. The result is
\bea
\tau_{\omega}(L_{2N-1}(g)) = \left|\prod_{i=1}^N \(2 \sin \frac{\pi q_i^{-1} \ell}{p}\)\right|
\eea
where we define $q_i^{-1}$ as an integer such that
\ben
q_i^{-1} q_i &\equiv & 1\label{mod}
\een
mod $p$.\footnote{The absolute value seems unnatural. In Ray's original computation \cite{Ray}, he computed the square of what we call the Ray-Singer torsion here. The square is real and positive and has no sign ambiguity under $\l\rightarrow \l + p$. In that sense it is the squared object that is the natural object to consider. But here we follow the widely used custom, and define the Ray-Singer torsion as the positive square root of Ray's original definition of the torsion. When $N$ is even, the absolute value is not necessary. However, in this paper we will be considering both cases when $N$ is even and odd, so we need to have absolute value.}

The condition (\ref{mod}) is also valid for $N=1$, which is a circle with a twisted boundary condition. We start with the lens space $\t L(p;1)$, and twisted boundary condition
\ben
\phi(e^{\frac{2\pi i}{p}} z) &=& e^{\frac{2\pi i \l'}{p}} \phi(z)\label{bc}
\een
for some $\l'\neq 0$. By an explicit computation using the Hurwitz zeta function regularization, one can find the torsion
\bea
\tau_{\l'}(\t L(p;1)) &=& \left|2 \sin \frac{\pi \l'}{p}\right|
\eea
For the lens space $\t L(p;q_1)$ we shall consider the twisted boundary condition
\bea
\phi(e^{\frac{2\pi i q_1}{p}} z) &=& e^{\frac{2\pi i \l}{p}} \phi(z)
\eea
for some $\l$. We get this boundary condition by iterating (\ref{bc}) $q_1$ times, which gives $\l = q_1 \l'$ and the torsion can be written in the form
\bea
\tau_{\l}(\t L(p;q_1)) = \left|2 \sin \frac{\pi \l'}{p}\right| = \left|2 \sin \frac{\pi q_1^{-1} \l}{p}\right|
\eea
This extends Ray's computation, which is valid for $N\geq 2$, to the case of $N=1$.

For a general lens space, we can always fix $q_N=1$ without imposing any restrictions. We then use the notation $L(p;q_1,\dots q_{N-1}) := \t L(p;q_1,\dots,q_{N-1},1)$ for the lens space. 

We can also compute the Ray-Singer torsion with a trivial holonomy $\l=0$. In this case the result is  
\ben
\tau_{0} &=& \frac{1}{p^{N-1}}\label{general}
\een
We show this result by explicit computations in appendix \ref{RSE} for $N=1$ and $N=3$. The case of $N=2$ was addressed in the appendix of the paper \cite{Friedmann:2002ty}. From these results, we conjecture that the above formula will hold for all integers $N=1,2,3,...$.

Let us now obtain the Ray-Singer torsion for $p$-forms taking values in the fundamental representation of the gauge group $U(N)$. The lens space $L(p;q_1)$ has the fundamental group $\mb{Z}_p$ and the holonomies are maps from the fundamental group into $U(N)$, labeled by integers $\l_i=0,\dots, p-1$ for $i=1,\dots, N$. Alternatively we can consider a partition of $N = N_0 + N_1 + \cdots + N_{p-1}$ where $N_{\l}$, for $\l=0,\dots, p-1$, counts the number of indices $i=1,\dots, N$ for which $\l_i = \l$. The twisted boundary conditions are such that the field component $\phi_{ij}$ carries charge $\l_i-\l_j$ under the $U(1)$ that rotates along the Hopf fiber of $L(p;q_1)$. The Ray-Singer torsion for $U(N)$ is given by the product of the torsions for all the field components $ij$. This can be expressed as
\bea
\tau_{(\l_1,\dots, \l_N),U(N)} = \[\tau_{0}\]^{N'} \prod_{\l\neq \l'}^{p-1} \[\tau_{\l-\l'}\]^{N_{\l} N_{\l'}} 
\eea
where $N' = \sum_{\l=0}^{p-1} N_{\l}^2 \geq N$. In the generic situation where all the $\l_i$'s are distinct so that each $N_{\l} \leq 1$, we have $N'=N$. To get the torsion for $SU(N)$ gauge group, we impose the restriction $\l_1+\cdots + \l_N = 0$ mod $p$. For $SU(2)$ we get $N_{\l} = N_{p-\l} = 1$ for some $\l$, and all other $N_{\l'} = 0$. The Ray-Singer torsion then becomes
\bea
\tau_{\l,SU(2)} = \tau_{0} \tau_{2\l} \tau_{-2\l} &=& \frac{16}{p} \(\sin \frac{2\pi q_1^{-1} \l}{p} \sin \frac{2\pi \l}{p}\)^2
\eea

We will now construct flat gauge fields with nontrivial holonomies for $L(p;q)$ and for $L(p;q_1,q_2)$. We begin with considering the orbifold $\mb{C}^2/\mb{Z}_p$ with the orbifold identification
\bea
\(z_1,z_2\) &\sim &\(z_1 e^{\frac{2\pi i q}{p}},z_2 e^{\frac{2\pi i}{p}}\)
\eea
The lens space $L(p;q)$ is defined by the equation $|z_1|^2+|z_2|^2 = 1$ in this orbifold. But the lens space is smooth since the orbifold singularity is at the origin, away from the lens space. Let us consider two coordinate patches, $U_1 = \{z_1 \neq 0\} = \{0<r\}$ and $U_2 = \{z_2 \neq 0\} = \{r< 1\}$ with $r=|z_1|$. On the patch $U_2$, we define
\bea
z^1 &=& \sin\frac{\theta}{2}e^{i\frac{q}{p} \psi_2 - i \phi_2}\cr
z^2 &=& \cos\frac{\theta}{2}e^{i\frac{1}{p} \psi_2}
\eea
We have the following rectangular $T^2$-identifications
\bea
\psi_2 &\sim & \psi_2 + 2\pi\cr
\phi_2 &\sim & \phi_2 + 2\pi
\eea
Of course the metric on this $T^2$ is complicated, induced from the flat metric on $\mb{C}^2$, but here the metric is not our concern. The above $T^2$ identifications are preserved by the mapping class group $SL(2,\mb{Z})$. To go to the patch $U_1$ we shall preserve the $T^2$ identifications, and hence the coordinate transformation must correspond to some element of $SL(2,\mb{Z})$. Indeed this is the case. The transformation that does the job reads
\bea
\psi_2 &=& m \psi_1 + p \phi_1\cr
\phi_2 &=& n \psi_1 + q \phi_1
\eea
where $m$ and $n$ are chosen so that 
\ben
mq - np &=& 1\label{mn3d}
\een
The existence of such $m$ and $n$ follows from Bezout's theorem and the assumption that $p$ and $q$ are relatively prime. 

We have the same torus identifications after the $SL(2,\mb{Z})$ transformation,
\bea
\psi_1 &\sim & \psi_1 + 2\pi\cr
\phi_1 &\sim & \phi_1 + 2\pi
\eea
We get
\bea
z^1 &=& \sin\frac{\theta}{2}e^{i\frac{1}{p} \psi_1}\cr
z^2 &=& \cos\frac{\theta}{2}e^{i\frac{m}{p} \psi_1 + i\phi_1}
\eea
which are coordinates on the patch $U_1$. In particular we notice that since $mq \equiv 1$ mod $p$, we realize the lens space identification by taking $\psi_1 \rightarrow \psi_1 + 2\pi q$. Further since $p$ and $q$ are relative prime, we generate all elements of $\mb{Z}_p$ by taking integer multiples of $q$ if we count modulo $p$.

We now seek a flat connection which corresponds to the holonomy
\bea
\exp \frac{2\pi i \ell}{p}
\eea
when integrated over the closed path $C$ that is specified by (for $t\in[0,2\pi]$)
\bea
\begin{pmatrix}
\psi_1\\
\phi_1
\end{pmatrix}
&=& \begin{pmatrix}
t\\
0
\end{pmatrix}
\eea
using the coordinates on $U_1$, and by 
\bea
\begin{pmatrix}
\psi_2\\
\phi_2
\end{pmatrix}
&=& \begin{pmatrix}
m t\\
n t
\end{pmatrix}
\eea
using the coordinates on $U_2$.

One such flat connection on the patch $U_1$ is given by
\bea
A|_{U_1} &=& \frac{\ell}{p} d \psi_1 
\eea 
which is well defined over the entire $U_1$. Note that over $U_1$, the $T^2$ collapses   to a circle when $|z_2|=0$ or $r=1$ and, even in the region near the circle, this connection $A_1$ is well defined. In the overlap region $U_1\cap U_2$, the coordinate transformed version of $A|_{U_1}$ is given by 
\bea
A|_{U_1\cap U_2} &=& \frac{\ell}{p} \(q d\psi_2 - p d \phi_2\)
\eea
Again note that over $U_2$, $T^2$ collapses to a circle when $|z_1|=r=0$.  To make $A$ well defined all over $U_2$, we need to make a large gauge transformation 
\bea
A' = A + \ell d \phi_2 =  \frac{q \ell}{p} d\psi_2 
\eea 
in the overlap of $U_1\cap U_2$. Then the connection 
\bea
A'|_{U_2} &=& \frac{q\l}{p} d\psi_2
\eea
is well-defined all over $U_2$. 

Using (\ref{mn3d}) it is easy to see that  
\bea
\exp i \int_C A = \exp i \int_C A' = \exp \frac{2\pi i \l}{p}
\eea
if $C$ is in $U_1\cap U_2$. If $C$ is not in $U_1$, then we compute the holonomy using the expression $\exp i \int_C A'$ and if $C$ is not in $U_2$, then we use $\exp i \int_C A$ and we get the same result. Thus we have shown that there exists a flat gauge field on $L(p;q)$ with the above given holonomy around $C$.

In the fundamental representation of $SU(2)$ gauge group we wish to find a flat gauge field corresponding to the holonomy
\bea
U_{\ell} &=& \begin{pmatrix}
\omega^{\ell} & 0\\
0 & \omega^{p-\ell}
\end{pmatrix}
\eea
where we define $\omega = e^{2\pi i/p}$. In the path integral, we shall sum over all gauge inequivalent holonomies. Since the Weyl group, which is a subgroup of the gauge group, permutes the two diagonal elements in the holonomy, we see that gauge inequivalent holonomies are obtained by restricting the range to $\ell = 0,1,...,[p/2]$ where $[p/2]=p/2$ if $p$ is even, and $(p-1)/2$ if $p$ is odd. If $p$ is even, then we find two elements in the center, $U_0 = diag(1,1)$ and $U_{p/2}=(-1,-1)$. If $p$ is odd, the only element in the center is $U_0$.

A corresponding flat gauge field in the fundamental representation of $SU(2)$ is given by
\bea
A|_{U_1} &=& \frac{\l}{p} \begin{pmatrix}
1 & 0\\
0 & -1
\end{pmatrix} d\psi_1 
\eea

For $U(1)$ gauge group, flat gauge fields on the lens space $L(p;q)$ are classified by the holonomy around the Hopf fiber 
\bea
\exp i \int_C A &=& \exp \frac{2\pi i \ell}{p}
\eea
where $\ell =0,...,p-1$. These take values in $\pi_1(L(p;q)) = \mb{Z}_p$. The Chern-Simons action is given by
\bea
S &=& -\frac{ik}{4\pi} \int_{L(p;q)} A \wedge dA
\eea
in Euclidean signature. This action has been explicitly computed for $q=p-1$ on a flat gauge field in \cite{Imamura:2013qxa}. The result is\footnote{Since we are in Euclidean signature, we need an $i$ in the CS action. The minus sign in the right hand side is related to the choice of orientation of the lens space.} 
\bea
\exp(-S) &=& \exp\(-\pi i k \frac{1-p}{p} \ell^2\)
\eea
This result is nontrivial, despite $F=dA=0$ and so the CS action would naively be zero. But the above definition of the CS action is not entirely correct when we need to cover the manifold with many patches and the gauge field is related by gauge transformations as we pass from one patch to another. A better way to define the CS action in such a situation, is as $S \sim \int_{M_4} F \wedge F$ where $\partial M_4 = L(p;q)$. Here $F$ does not have to vanish on $M_4$. For the detailed construction of $F$, we refer to \cite{Imamura:2013qxa}.

For $SU(2)$ gauge group the value of Chern-Simons action has been obtained for a flat gauge field corresponding to the holonomy 
\bea
P\exp i \int_C A &=& \begin{pmatrix}
\exp \frac{2\pi i \ell}{p} & 0\\
0 & \exp -\frac{2\pi i \ell}{p}
\end{pmatrix}
\eea
for all lens spaces in \cite{Kirk}. The result is 
\bea
\exp(-S) &=& \exp\(\frac{2\pi i k q^{-1} \ell^2}{p}\)
\eea
where $q^{-1}q = 1$ mod $p$. For $q=p-1$ we have $q^{-1} =p-1$ and
\bea
\exp(-S) = \exp\(2\pi i k \frac{p-1}{p} \ell^2\) = \exp\(-\frac{2\pi i k\ell^2}{p}\)
\eea
There is a factor $2$ compared to the $U(1)$ case, which we can understand as follows. The $SU(2)$ CS action on this flat gauge field with the above holonomy is given by
\bea
A dA + (-A)d(-A) = 2 A dA
\eea
where $A$ denotes the $U(1)$ gauge field. We may also notice that the minus sign in the second step is consistent with taking $k$ to $-k$ in the resulting CS action for the lens space $L(p;1)$. The lens spaces $L(p;1)$ and $L(p;p-1)$ are related by parity, which flips the sign of $k$.

We now move on the $L(p;q_1,q_2)$. We view $S^5$ as a circle bundle over $\mb{CP}^2$ that we can cover with three patches 
\bea
U_a &=& \{(z_1,z_2,z_3)\in S^5 \subset \mb{C}^3 | z_a \neq 0\}
\eea
The lens space $L(p;q_1,q_2)$ is defined as $S^5/\mb{Z}_p$ where
\ben
\(z_1,z_2,z_3\) &\sim &\(z_1 e^{\frac{2\pi i q_1}{p}},z_2 e^{\frac{2\pi i q_2}{p}},z_3 e^{\frac{2\pi i}{p}}\)\label{lens5d}
\een
On $U_3$ we use the coordinates 
\bea
z_1 &=& \sin\chi \cos\frac{\theta}{2} e^{i \frac{q_1}{p} y_3 + i \psi_3 + i \phi_3}\cr
z_2 &=& \sin\chi \sin\frac{\theta}{2} e^{i \frac{q_2}{p} y_3 - i \psi_3}\cr
z_3 &=& \cos\chi e^{i \frac{1}{p} y_3}
\eea
The lens space identification (\ref{lens5d}) is obtained by taking $y_3 \rightarrow y_3 + 2\pi$. The coordinates $\chi$ and $\theta$ may be used on all three patches if we understand that their ranges are different depending on the patch,
\bea
U_3 &=& \{0\leq \chi < \pi/2,\quad 0\leq \theta \leq \pi\}\cr
U_2 &=& \{0<\chi\leq \pi/2,\quad 0<\theta\leq \pi\}\cr
U_1 &=& \{0<\chi\leq \pi/2,\quad 0\leq \theta <\pi\}
\eea
The coordinates $(y_3,\psi_3,\phi_3)$ take values in a three-torus $T^3 = \mb{R}^3 / (2\pi \mb{Z})^3$. We map from $U_3$ to $U_2$ by the following $SL(3,\mb{Z})$ coordinate transformation
\bea
\begin{pmatrix} 
y_3 \\
\psi_3 \\
\phi_3
\end{pmatrix} &=& \begin{pmatrix}
m & p & 0\\
n & q_2 & 0\\
-n & 1-q_1-q_2 & 1
\end{pmatrix}
\begin{pmatrix}
y_2\\
\psi_2\\
\phi_2
\end{pmatrix}
\eea
where $m$ and $n$ are such that 
\ben
m q_2 - n p &=& 1\label{mn}
\een
Since $SL(3,\mb{Z})$ is the mapping class group of $T^3$, we have $(y_2,\phi_2,\phi_2)\in T^3$ and
\bea
z_1 &=& \sin\chi \cos\frac{\theta}{2} e^{i \frac{q_1m}{p}y_2+i\psi_2+i\phi_2}\cr
z_2 &=& \sin\chi \sin\frac{\theta}{2} e^{i \frac{1}{p} y_2}\cr
z_3 &=& \cos\chi e^{i \frac{m}{p} y_2 + i \psi_2}
\eea
and the lens space identification (\ref{lens5d}) is obtained by taking $y_2 \rightarrow y_2 + 2\pi q_2$. 

We map from $U_3$ to $U_1$ by the following $SL(3,\mb{Z})$ transformation,
\bea
\begin{pmatrix} 
y_3 \\
\psi_3 \\
\phi_3
\end{pmatrix} &=& \begin{pmatrix}
\t m & -p & -p\\
0 & 1-q_2 & -q_2\\
-\t n & -1+q_1+q_2 & q_1+q_2
\end{pmatrix}
\begin{pmatrix}
y_1\\
\psi_1\\
\phi_1
\end{pmatrix}
\eea
where $\t m$ and $\t n$ are such that
\ben
\t m q_1 - \t n p &=& 1\label{mn1}
\een
Then we get 
\bea
z_1 &=& \sin\chi\cos\frac{\theta}{2} e^{i\frac{1}{p} y_1}\cr
z_2 &=& \sin\chi\sin\frac{\theta}{2} e^{i\frac{q_2 \t m}{p} y_1 - i \psi_1}\cr
z_3 &=& \cos\chi e^{i\frac{\t m}{p}y_1 - i\psi_1 - i\phi_1}
\eea
By using $q_1 \t m \equiv 1$ mod $p$, we find the lens space identification (\ref{lens5d}) by taking $y_1 \rightarrow y_1 + 2\pi q_1$.

The map from $U_2$ to $U_1$ can be obtained by composing the map from $U_2$ to $U_3$ (the inverse of the map from $U_3$ to $U_2$) and the map from $U_3$ to $U_1$. 

In the overlap $U_1\cap U_2\cap U_3$, we take the gauge field as
\bea
A &=& \frac{\ell}{p} dy_3= \frac{\ell}{p} \(m dy_2 + p d\psi_2\) = \frac{\ell}{p} \(\t m dy_1 - p d\psi_1 - p d\phi_1\)
\eea
Just as we did in 3d case, here again we shall remove Dirac string singularities in order to have a well-defined gauge potential on the patches $U_2$ and $U_3$. Thus we define
\bea
A|_{U_3} &=& \frac{\l}{p} dy_3\cr
A'|_{U_2} &=& \frac{\l}{p} m dy_2\cr
A''|_{U_1} &=& \frac{\l}{p} \t m dy_1
\eea
On the overlap regions these gauge fields are related by large gauge transformations. The path $C$ along which we integrate the holonomy is expressed as follows in the three coordinate patches respectively as follows,
\bea
(y_3,\psi_3,\phi_3) &=& (1,0,0)t\cr
(y_2,\psi_2,\phi_2) &=& (q_2,-n,(1-q_1)n)t\cr
(y_1,\psi_1,\phi_1) &=& (q_1,q_2\t n,(1-q_2)\t n)t
\eea
where $t\in[0,2\pi]$. The holonomy remains the same after the large gauge transformations and is the same irrespectively of which patch we compute 
it in and is given by
\bea
\exp i \int_C A = \exp i \int_C A' = \exp i \int_C A'' = \exp \frac{2\pi i \l}{p}
\eea
To see this, we use the relations (\ref{mn}) and (\ref{mn1}).

For $SU(2)$ gauge group we also have a flat gauge field that we get simply multiplying the $U(1)$ flat gauge field by the matrix diag$(1,-1)$ and for $U(N)$ gauge group we have the flat gauge field 
\bea
A|_{U_3} &=& \frac{1}
{p}
 \begin{pmatrix}
\l_1 & & \\
      & \ddots & \\
      &    &  \l_N
\end{pmatrix}
dy_3
\eea
Thus we find that to each possible holonomy
\bea
\exp \frac{2\pi i}{p} \begin{pmatrix}
\l_1 & & \\
      & \ddots & \\
      &  & \l_N
      \end{pmatrix}
      \eea
around the fiber of $L(p;q_1,q_2)$, there is a corresponding flat gauge field, which is defined on each coordinate patch and related to different patches by gauge transformations.

\section{One-dimensional fermionic Chern-Simons}
Before turning to the more complicated case of 5d FCS, we will consider 1d FCS, that is, quantum mechanics with one real fermion. Let us consider a hermitian operator $a$ subject to the algebra
\bea
a^2 &=& \1
\eea
and assume that there is one state $\|0\>$. Acting with $a$ we get another state $\|1\> = a \|0\>$. It is not possible for $a$ to annihilate $\|0\>$, since by acting twice by $a$ we shall get back $\|0\>$. If we act by $a$ on $\|1\>$ we get $a\|1\> = \|0\>$ by using $a^2 = 1$. Let us normalize the state as $\Braket{0|0} = 1$. Inserting $\1 = a^{\dag} a$, we get $\Braket{1|1} = 1$. We have $\Braket{0|1} = \Braket{1|0} = 0$ as a consequence of the requirement that the $2\times 2$ matrix realization of $a$ shall square to the identity matrix. We have the completeness relation
\bea
\|0\>\<0\! + \|1\>\<1\! &=& \1
\eea
We introduce a Grassmann odd parameter $\psi$ such that $\psi \psi = 0$ which we can integrate over with the usual rules 
\bea
\int d\psi &=& 0\cr
\int d\psi \psi &=& 1
\eea
We have the anticommutativity property,
\bea
a \psi &=& -\psi a
\eea
We define the state
\bea
\|\psi\> &=& \|0\> + \|1\> \psi
\eea
The conjugate state is 
\bea
\<\psi\! &=& \<0\! + \psi \<1\! 
\eea
We have the following properties
\bea
\Braket{\psi'|\psi} &=& e^{\psi'\psi}\cr
\Braket{\psi'|a|\psi} &=& \psi' + \psi\cr
\int d\psi \|-\psi\> \<\psi\!a &=& \1\cr
\int d\psi \<\psi\!a A\|\psi\> &=& \tr A
\eea
The partition function can be written as
\bea
Z &=& \tr \1 = \int d\psi \<\psi\!a\|\psi\> = \int d\psi \<\psi\!a (aa) (aa) \cdots (aa)\|\psi\> \cr
&=& \int d\psi \<\psi\!\1 a (\1 a\1 a) \cdots (\1 a\1 a)\|\psi\>
\eea
We use the completeness relation which absorbs all the operators $a$ and we get
\bea
Z &=& \int d\psi d\psi_N \cdots d\psi_1 \Braket{\psi|-\psi_N} \Braket{\psi_N|-\psi_{N-1}} \cdots \Braket{\psi_1|\psi}\cr
&=& \int d\psi d\psi_N \cdots d\psi_1 e^{-\psi \psi_N + \psi_N \psi_{N-1} + \cdots + \psi_1 \psi}\cr
&=& \int d\psi d\psi_N \cdots d\psi_1 e^{-\(\epsilon (-\psi) \frac{(-\psi)-\psi_N}{\epsilon} + \cdots + \epsilon \psi_1 \frac{\psi_1 - \psi}{\epsilon}\)}
\eea
We can compute this partition function for any odd integer $N$ and always get the same answer,
\bea
Z &=& 2
\eea
which counts the number of states. In the limit $N\rightarrow \infty$, we get the path integral 
\bea
Z &=& \int \D\psi e^{iS}
\eea
where we take the antiperiodic boundary condition $\psi(2\pi) = -\psi(0)$ and the FCS action is given by
\bea
S &=& \frac{k}{4\pi} \int_0^{2\pi} dt \psi \dot\psi
\eea
Here $k$ is a real parameter that will be determined by matching with canonical quantization. Since the fermionic fields are anticommuting, the FCS action is purely imaginary. We find that this choice is necessary in order for the partition function to become a real number. We compute the path integral by expanding the field in orthonormalized modes with respect to the metric (or line element) $dt$. Our mode expansion reads
\bea
\psi(t) &=& \sum_{n\in \mb{Z}} \psi_{n+\frac{1}{2}} e^{i\(n+\frac{1}{2}\)t}
\eea
The action is 
\bea
S &=& - ik \sum_{n=0}^{\infty} \(n+\frac{1}{2}\)\psi_{n+\frac{1}{2}}\psi_{-n-\frac{1}{2}}
\eea
We compute this using Hurwitz zeta function as follows,
\bea
Z &=& \prod_{n=0}^{\infty} \[k \(n+\frac{1}{2}\)\] = \sqrt{k} e^{-\zeta'(0,\frac{1}{2})} = \sqrt{2k}
\eea
where we choose the fermionic measure as
\bea
\D \psi = \prod_{n=0}^{\infty} d\psi_{n+\frac{1}{2}} d\psi_{-n-\frac{1}{2}}
\eea
To match with the result from canonical quantization, that is $Z = 2$, we shall take $k=2$. 

Let us next compute the Witten index. We define a fermion number operator as
\bea
(-1)^F\|0\> &=& \|0\>\cr
(-1)^F\|1\> &=& -\|1\>
\eea
This means that 
\bea
(-1)^F a &=& -a (-1)^F\cr
(-1)^F \|\psi\> &=& \|-\psi\>
\eea
The Witten index is
\bea 
I = \tr(-1)^F = \int d\psi \<\psi\! a (-1)^F \|\psi\> = -\int d\psi \<-\psi\! a \|\psi\>
\eea
If we then follow through the same steps as above, we end up with the same path integral but with the boundary condition $\psi(2\pi) = \psi(0)$ and in that case we get 
\bea
I &=& 0
\eea
when we compute the resulting expression for any odd number $N$ of steps. This is easy to see for $N=1$. Then the exponent becomes $\psi\psi_1+\psi_1\psi = 0$ and the integration over $\psi$ and $\psi_1$ yields zero.

Since $a^2 = \1$, we can introduce projection operators
\bea
P_{\pm} &=& \frac{1}{2} \(1\pm a\)
\eea
Orthonormalized eigenstates of $P_{\pm}$ are
\bea
\|\pm\> &=& \frac{1}{\sqrt{2}} \(\|0\> \pm \|1\>\)
\eea
Now $a\|\pm\>=\pm\|\pm\>$, and $(-1)^F$ maps the two states into each other, $(-1)^F \|\pm\> = \|\mp\>$. We have the projected Witten index  
\bea
I_+ = \tr \((-1)^F P_+\) = \<+\!(-1)^F\|+\>= \Braket{+|-} = 0
\eea
We can imagine a different definition where we instead define $(-1)^F = a$. With this definition we declare $\|+\>$ is bosonic and $\|-\>$ is fermionic and then we get
\bea
I_+ = \tr \((-1)^F P_+\) = \<+\!a\|+\>= \Braket{+|+} = 1
\eea
To connect with the path integral, an attempt would be to define
\bea
\|\psi\> &=& e^{\psi} \|+\>
\eea
since then, by using $a \psi = - \psi a$, 
\bea
(-1)^F \|\psi\> &=& \|-\psi\>
\eea
which leads to a path integral with periodic boundary condition. However, the action turns out to become nonlocal. 

Let us next compute the Witten index by taking out the zero mode. We then consider the path integral 
\bea
I &=& \int \D\psi e^{i S}
\eea
with periodic boundary condition. We expand the field in modes
\bea
\psi(t) &=& \sum_{n\neq 0} \psi_n e^{int}
\eea
The action becomes
\bea
S &=& -ik \sum_{n=1}^{\infty} n \psi_n \psi_{-n}
\eea
The index becomes
\bea
I_{osc} = \prod_{n=1}^{\infty} kn = \sqrt{\frac{2\pi}{k}}
\eea
where we use the fermionic measure 
\bea
\D \psi = \prod_{n=1}^{\infty} d\psi_{n} d\psi_{-n}
\eea
Now we have gauge fixed a fermionic zero mode. As we illustrate in the appendix \ref{gf1}, when we do this, we shall also divide the result by the volume of the gauge group. Here the gauge group depends on the context. If the gauge group is trivial, we shall not divide by anything. But if the gauge group is $U(1)$ acting trivially on $\psi$ in the adjoint representation, then we shall divide by its volume $V$. 

Let us now consider the following nonabelian generalization with the action
\bea
S &=& \frac{k}{8\pi} \int_0^{2\pi} dt \tr_R(\psi D_t \psi)
\eea
where the trace is taken in some representation $R$ of the gauge group. We define $D_t \psi = \dot{\psi} - i [A_t,\psi]$ where $A_t$ is a background gauge field. Let us now restrict to the gauge group $SU(2)$ and let the gauge field be $A_t = A_{t3} T^3$ where $[T^3,T^{\pm}] = \pm T^{\pm}$. We expand $\psi = \psi_3 T^3 + \psi_+ T^+ +\psi_- T^-$ and find
\bea
D_t \psi_3 &=& \dot\psi_3\cr
D_t \psi_{\pm} &=& \dot\psi_{\pm} \mp i A_{t3} \psi_{\pm}
\eea
Let us assume the generators are in the fundamental representation, $R=\Box$, of $SU(2)$ and let us normalize the generators so that $\tr_\Box(T^3 T^3) = 2$ and $\tr_\Box(T^+ T^-) = 1$. Then the action becomes a sum of two terms, $S = S_3 + S_+$ where
\bea
S_3 &=& \frac{k}{4\pi} \int_0^{2\pi} dt \psi_3 \dot\psi_3 \cr
S_+ &=& \frac{k}{4\pi} \int_0^{2\pi} dt \(\psi_+ \dot\psi_- - i A_{t3} \psi_+ \psi_-\)
\eea 
The holonomy is 
\bea
U &=& \begin{pmatrix} e^{i h} & 0\\ 0 & e^{-i h} \end{pmatrix}
\eea
where $h = \int_0^{2\pi} dt A_{t3}$. The Witten index splits into a product $I = I_3 I_+$. From the previous result we get from $S_3$ the contribution $I_3 = I_{osc}/V$. From $S_+$ we get 
\bea
I_+ = \prod_{n\in \mb{Z}} \(n+ \frac{h}{2\pi}\) = 
2i  \sin \(\frac{h}{2}\)
\eea
for $h\neq 0$, where  in the last step we used zeta function regularization. This does not depend on $k$. We also know that the Witten index shall be $ \pm
 2 i \sin (h/2)$ from canonical quantization, since this is the sum over two states, spin-down and spin-down, with a relative minus sign as we are computing a Witten index. This gives the sine function multiplied by the factor of $\pm 2i$. The sign depends on how we define the fermion number $F$ when we define the Witten index -- whether the spin-up state is taken as the fermionic or the bosonic state.

For general gauge group, holonomy $U\neq \1$ in the representation $R$, the Ray-Singer torsion on $S^1$ is given by \cite{Witten:1991we,Blau:1993tv}
\bea
\tau &=& \det{}_R (\1 - U)
\eea
If we specialize this formula to the fundamental representation of $SU(2)$, then this yields
\bea
\tau = (1 - e^{i h}) (1 - e^{-i h}) = \(2\sin\(\frac{h}{2}\)\)^2
\eea
We see that up to a phase factor $\pm i$, the fermionic Chern-Simons action computes the square root of the RS torsion on $S^1$. 

We can restrict ourselves to $h = 2\pi (2\l)/p$ in which case we can view this as the torsion of the lens space $L(p)=S^1/\mb{Z}_p$. The partition function is given by 
\ben
I &=& I_3 I_+(0) + \sum_{\l=1}^{\frac{p-1}{2}} I_3 I_+(\l) \label{a sum}
\een
where
\bea
I_3 &=& \frac{I_{osc}}{V} \cr
I_+(0) &=& (I_{osc})^2\cr
I_+(\l) &=& 2 i \sin \frac{2\pi \l}{p}
\eea
This expression is explained as follows: when $\l = 0$ we get the zero mode contribution from $\psi_3, \psi_+, \psi_-$, whereas when $\l\neq 0$ we get the zero mode contribution only from $\psi_3$. We note that $U(1)$ inside $SU(2)$ is generated by $T^3$, so for $I_3$ we divide by $V$ but for $I_+(0)$ we do not divide by $V$ since there is no $U(1)$ associated with $I_+(0)$. We get the squared expression $I_+=(I_{osc})^2$ as can be understood by rewriting the Lagrangian in terms of real spinor components: $\psi_+\dot\psi_- = \psi_1\dot\psi_1+\psi_2\dot\psi_2$. Now we can write this in the form
\ben
I &=& e^{\frac{3 \pi i}{2}} \[\frac{1}{\Vol(SU(2))} \sqrt{\tau_{0,SU(2)}} + \frac{1}{\Vol(U(1))} \sum_{\l=1}^{\frac{p-1}{2}} \sqrt{\tau_{\l,SU(2)}}\]\label{Z1dFCS}
\een
where the radius of $SU(2)$ and the volume $V$ shall be taken as
\ben
r &=& \sqrt{\frac{i}{\pi}} \sqrt{\frac{k}{2\pi}},\label{radiusFCS}\\
V &=& 2\sqrt{\pi}
\een
To show this result, we need to notice that $I_+(\l)$ lies in the upper halfplane for all values $\l=1,...,(p-1)/2$. The overall phase factor $e^{\frac{3 \pi i}{2}}$ is should somehow come from the BRST gauge fixing of the fermionic zero mode, since that gives the same factor for all the holonomy sectors. We notice that by requiring $V$ coincides with $\Vol(U(1)) = 2\pi r$ we have to fix $k=-2\pi i$. For this value the abelian FCS action becomes
\bea
S &=& -\frac{i}{2} \int_0^{2\pi} dt \psi \dot\psi
\eea
which is both real and canonically normalized.

\section{Five-dimensional fermionic Chern-Simons}
In \cite{Bak:2015hba} we found that maximally twisted 5d MSYM gives a 5d fermionic Chern-Simons theory. This is a topological field theory. We are now interested in its partition function. Since the action is topological, the partition function should be a topological invariant of the five-manifold $M_5$.

Let us begin with assuming the gauge group is $U(1)$. If we introduce the linear combinations
\bea
\A_m &=& A_m - \phi_m\cr
\b\A_m &=& A_m + \phi_m
\eea
then the Lagrangian can be expressed as 
\bea
\L = \L_{FCS} + \delta V
\eea
where the fermionic Chern-Simons term is
\bea
\L_{FCS} &=& - \frac{i}{8} \epsilon^{mnpqr} \psi_{mn} \partial_p \psi_{qr}
\eea
and the gauge fixing fermion that arises from twisting of 5d MSYM is given by
\bea
V &=& \frac{1}{4} \b{\F}_{mn}\psi^{mn}-\frac{1}{2}\phi\psi-\phi^m\partial_m\psi
\eea
By an integration by parts we can write this as
\bea
V &=& -\frac{1}{2} \b{\A}_n \b{\D}_m \psi^{mn} - \frac{1}{2}\phi\psi-\phi^m\partial_m\psi
\eea
which is on the form that shows that this will correspond to the gauge fixing condition
\bea
\b{\D}_m \psi^{mn} &=& 0
\eea
The BRST variations, which are inherited from the 5d MSYM supersymmetry upon twisting, read 
\bea
\delta \psi_{mn} &=& \F_{mn}\cr
\delta \psi_m &=& 0\cr
\delta \psi &=& - \phi\cr
\delta \b\A_m &=& -2i \psi_m\cr
\delta \A_m &=& 0\cr
\delta \phi &=& 0
\eea
While these fix the two-form gauge symmetry for $\psi_{mn}$, they still leave a residual gauge symmetry for $A_m$ which we need to further BRST gauge fix. We do that in the usual fashion by adding the anticommuting $c$ and $\bar{c}$ ghosts and the auxiliary field $B$ for which we have the standard Yang-Mills BRST variations
\bea
\delta' A_m &=& \partial_m c\cr
\delta' B &=& 0\cr
\delta' c &=& 0\cr
\delta' \bar{c} &=& i B
\eea
These are nilpotent, ${\delta'}^2 = 0$ and they commute with the supersymmetry variations, $\{\delta,\delta'\} = 0$. A convenient choice for this second gauge fixing fermion is
\bea
V' &=& -i \bar{c}\(\nabla^m \A_m + \phi - \frac{\alpha}{2} B\)
\eea
We then get
\bea
\delta' V' &=& B \(\nabla^m \A_m + \phi\) - \frac{\alpha}{2} B^2 + i \bar{c} \nabla^m \partial_m c\cr
\delta V &=& \frac{1}{4}\b\F^{mn}\F_{mn} + \frac{1}{2} \phi^2 + \phi^m \partial_m \phi \cr
&&+ \frac{i}{2} \psi^{mn} \(\partial_m \psi_n - \partial_n \psi_m\) + i \psi^m\partial_m \psi
\eea
and $\delta V' = \delta' V = 0$. The full Lagrangian is
\bea
\L &=& - \frac{i}{8} \epsilon^{mnpqr} \psi_{mn} \partial_p \psi_{qr} + \(\delta+\delta'\) \(V+V'\)
\eea
We now integrate out $\phi$ and then $B$. We then end up with 
\bea
\L &=& \frac{1}{4} F_{mn}^2 + \frac{1}{2(\alpha+1)} \(\nabla^m A_m\)^2\cr 
&&- \frac{1}{4} \phi_{mn}^2 - \frac{1}{2} \(\nabla^m \phi_m\)^2 + i \bar{c} \nabla^m \partial_m c\cr
&&+\frac{i}{2} \psi^{mn} \(\partial_m\psi_n-\partial_n\psi_m\) + i \psi^m \partial_m\psi\cr
&&- \frac{i}{8} \epsilon^{mnpqr} \psi_{mn} \partial_p \psi_{qr}
\eea
We then take $\alpha = 0$ to get the bosonic part of the action as
\bea
S_B &=& \frac{1}{2}(A,\triangle_1 A) - \frac{1}{2} (\phi,\triangle_1 \phi) + i (\bar{c},\triangle_0 c) 
\eea
The action for the fermions can be written in the form
\bea
S_F &=& \frac{1}{2} (\Psi,L\Psi)
\eea
where $\Psi := (\psi_2,\psi_1,\psi_0)^T$ and
\bea
L &=& i\(\begin{array}{ccc}
-*d & d & 0\\
-d^{\dag} & 0 & d\\
0 & -d^{\dag} & 0
\end{array}\)
\eea
is a hermitian operator that squares to 
\bea
L^2 &=& \(\begin{array}{ccc}
\triangle_2 & 0 & 0\\
0 & \triangle_1 & 0\\
0 & 0 & \triangle_0
\end{array}\)
\eea
We get the following contributions to the partition function,
\bea
Z_{\phi} &=& \frac{1}{\det{}^{\frac{1}{2}}\triangle_1}\cr
Z_{YM} &=& \frac{\det{}\triangle_0}{\det{}^{\frac{1}{2}}\triangle_1}\cr
Z_{F} &=& \det{}^{\frac{1}{4}} \triangle_2 \det{}^{\frac{1}{4}} \triangle_1  \det{}^{\frac{1}{4}} \triangle_0
\eea
coming from the fields $\phi_m$, the Yang-Mills gauge field $A_m$, and the fermionic fields $\psi,\psi_m,\psi_{mn}$ respectively. In these determinants, we can find zero modes. We will assume that $b_0 = b_5 = 1$ and that all the other Betti numbers are vanishing. This includes the lens spaces. In this case we have zero modes only coming from $\det\triangle_0$, which appears in both the fermionic part as well as the ghost part coming from the gauge fixing of the YM gauge potential. These are all fermionic ghost zero modes, which appear as a result of trying to gauge fix a gauge symmetry by a gauge fixing function that has a zero mode. These ghost zero modes come from fermionic ghosts associated with the gauge fixing of the fermionic two-form gauge symmetry and the YM gauge symmetry respectively. We should remove all ghost zero modes by further BRST gauge fixing. We present in detail how this is done in the appendix \ref{gf}. We define
\bea
Z_{osc} = Z_{\phi} Z'_{YM} Z'_F = \sqrt{\tau_{osc}}
\eea
where $\tau_{osc}$ is defined in (\ref{tauosc}) below and primes are used to indicate that zero modes are taken out from the determinants. The full partition function is obtained by dividing by the volume of the gauge group bundle $\G$
\ben
Z =  \frac{1}{\Vol(\G)} Z_{osc} \label{5dvol}
\een
In the appendix \ref{gf1} we argue that when we take out a fermionic ghost zero mode that is associated with a gauge fixing that leaves a residual gauge symmetry, we shall divide by a corresponding volume factor of the unbroken gauge group. See also \cite{Rozansky:1993zx}, \cite{Adams:1995xj}, \cite{Adams:1995np}, \cite{Adams:1996hi} and the lectures \cite{Marino:2011nm}. Now this result has been applied to nonabelian 3d CS perturbation theory where we expand around some flat background gauge field and the unbroken gauge group refers to the stability group for that background gauge field. But here we have an abelian gauge group and perhaps then the background field may be thought of as the gauge field is zero, around which we `expand' to quadratic order (that is, abelian theory with no interactions). That background gauge field being zero, does not break the abelian gauge group, so we have to divide by it. That is, we divide by the volume $\Vol(\G)$. Now the $Z_{osc}$ factor in the partition function has only a contribution coming from the oscillator modes since the zero modes have been taken out (or gauge fixed away). But we anticipate the full partition function will involve the Ray-Singer torsion which has both a zero mode part and an oscillator mode part. Indeed the zero mode contribution to the Ray-Singer torsion comes from the division by $\Vol(\G)$. As explained in \cite{Marino:2011nm} 
(Eq.~(3.64) in the arxiv version v5), we have 
\bea
\Vol(\G) &=& \Vol(U(1)) \Vol(M_5)^{\frac{1}{2}}
\eea
so we get
\bea
Z= \frac{1}{\Vol(U(1))} \sqrt{\tau}
\eea
where $\tau = \tau_{zero} \tau_{osc}$ is the Ray-Singer torsion, composed of the zero mode and the oscillator mode contributions
\ben
\tau_{zero} &=& \frac{1}{\Vol(M_5)}\\
\tau_{osc} &=& \frac{\det{}'^{\frac{1}{2}}\triangle_2\det{}'^{\frac{5}{2}}\triangle_0}{\det{}'^{\frac{3}{2}}\triangle_1}\label{tauosc}
\een
The Ray-Singer torsion is independent of the volume, but the determinants do depend on the volume. If $R$ denotes a typical length scale of $M_5$, then all the Laplacians will scale like $\triangle \sim R^{-2}$, and so 
\bea
\det \triangle \sim R^{-2\zeta_{\triangle}(0)} = R^{2 b_p}
\eea
where $b_p = \dim H_p$. Both steps in the above relation are nontrivial. We refer to appendix B for more details. Let us now consider the oscillator mode contribution to the Ray-Singer torsion,
\bea
\tau_{osc} = \prod_{p=0}^5 \(\det \triangle_p\)^{-(-1)^p \frac{p}{2}} \sim \prod_{p=0}^5 R^{- (-1)^p p b_p}
\eea
Thus this will have a nontrivial dependence on $R$, that we need to cancel by multiplying by a zero mode contribution. Let us assume that $b_0=b_5=1$ and all other Betti numbers are zero. Then 
\bea
\tau_{osc} \sim R^5
\eea
We thus need the zero mode contribution to be
\bea
\tau_{zero} \sim \frac{1}{\Vol(M_5)}
\eea
to cancel the dependence on $R$, as $\Vol(M_5) \sim R^5$. We verify explicitly this dependence on $R$ for the case that $M_5 = L(p;1,1)$ in the appendix \ref{RSE} where we get
\bea
\tau_{zero} &=& \frac{p}{\pi^3 R^5}\cr 
\tau_{osc} &=& \frac{\pi^3 R^5}{p^3}
\eea
The full partition function on $L(p;1,1)$ for abelian gauge group is now  
\ben
Z = \frac{p}{\Vol(U(1))} \sqrt{\tau} = \frac{1}{\Vol(U(1))}\label{abelian5dFCS}
\een
Here the factor of $p$ comes from summing over all holonomy sectors $\l = 0,...,p-1$. For abelian gauge group, the Weyl group is trivial and so we do not cut the sum at $\; = (p-1)/2$ as we do for $SU(2)$ gauge group. Since all fields are in adjoint which is trivial for $U(1)$ gauge group, all holonomy sectors give rise to the same result and we just sum them up which leads to a factor of $p$. The partition function is independent of $R$ and it is a topological invariant. 

\subsection{Nonabelian gauge group}
For a nonabelian gauge group with all the fields transforming in the adjoint representation we introduce the covariant derivatives
\bea
\D_m &=& \nabla_m - i \A_m\cr
\b\D_m &=& \nabla_m - i \b\A_m
\eea
The nonabelian fermionic Chern-Simons Lagrangian is given by 
\bea
\L_{5d} = \L_{FCS} + \delta V
\eea
where
\bea
\L_{FCS} &=& - \frac{i}{2} \epsilon^{mnpqr} \Tr \(\psi_{mn} \D_p \psi_{qr}\)
\eea
and the gauge fixing fermion is chosen as
\bea
V &=& \tr\(\frac{1}{2} \b{\F}_{mn}\psi^{mn}-\frac{1}{2}\phi\psi-\phi^m\D_m\psi\)
\eea
The 2-form BRST variations read 
\bea
\delta \psi_{mn} &=& \frac{1}{2} \F_{mn}\cr
\delta \psi_m &=& 0\cr
\delta \psi &=& - \phi\cr
\delta \b\A_m &=& -2i \psi_m\cr
\delta \A_m &=& 0\cr
\delta \phi &=& 0
\eea
As before these fix the two-form gauge symmetry for $\psi_{mn}$, and we need to fix the residual gauge symmetry for the one-form $A_m$. The standard Yang-Mills BRST variations would not commute with the above 2-form BRST variations, so instead we take these one-form BRST variations as
\bea
\delta' \b\A_m &=& 0\cr
\delta' \A_m &=& \D_m c\cr
\delta' B &=& 0\cr
\delta' c &=& \frac{i}{2}\{c,c\}\cr
\delta' \bar{c} &=& i B
\eea
These are still nilpotent, ${\delta'}^2 = 0$ but now these also commute with the 2-form BRST variations, $\{\delta,\delta'\} = 0$. A convenient choice for the second gauge fixing fermion is
\bea
V' &=& \tr\(-i \bar{c}\(\nabla^m \A_m + \phi - \frac{\alpha}{2} B\)\)
\eea
which gives 
\bea
\delta' V' &=& \tr\(B \(\nabla^m A_m - \nabla^m \phi_m + \phi\) - \frac{\alpha}{2} B^2 + i \bar{c} \nabla^m \D_m c\)\cr
\delta' V &=& 0\cr
\delta V' &=& 0\cr
\delta V &=& \tr\(\frac{1}{4} \b\F^{mn} \F_{mn} + \frac{1}{2} \phi^2 - \phi \D^m \phi_m\)
\eea
Then integrating out $\phi$ puts $\phi = \D^m \phi_m - B$ and then integrating out $B$ puts $B = \frac{1}{1+\alpha} \D^m \A_m$. Thus after integrating out all the auxiliary fields, we get
\bea
\L &=& \tr\(\frac{1}{4}\bar\F^{mn}\F_{mn}-\frac{1}{2}\(\D^m\phi_m\)^2 + \frac{1}{2(\alpha+1)} \(\D^m \A_m\)^2 - i \bar{c} \nabla^m \D_m c\)\cr
&&  -\frac{i}{8}\epsilon^{mnpqr} \tr\(\psi_{mn}\D_p\psi_{qr}\)
\eea

Preserving just one scalar real supercharge (or BRST charge), we can put the 5d FCS theory a generic five-manifold $M_5$, which has no isometries. Then supersymmetric field configurations satisfy
\bea
\F &=& 0\cr
d^{\dag}\phi &=& 0
\eea
Here\footnote{Here we Wick rotate $\phi_m$ into $i \phi_m$, which, as we explained in \cite{Bak:2015hba}, corresponds to Wick rotating time in the 6d theory. As we also explained there, BPS equations are better analyzed in this Euclidean theory.}
\bea
\F &=& F - i d\phi + i \phi^2
\eea
so that the BPS equations become
\bea
F + i \phi^2 &=& 0\cr
d\phi &=& 0\cr
d^{\dag}\phi &=& 0
\eea
This says that $\phi$ shall be a harmonic one-form. If we assume that the first cohomology group $H^1(M_5,\mb{Z})$ is trivial, this implies that 
\bea
F &=& 0\cr
\phi &=& 0
\eea
and the only BPS configurations are flat connections. 

Let us now assume that $M_5$ is a K-contact manifold. Then it has a unit normalizable Killing vector $v^m$. if $G_{mn}$ is the metric, then we define a contact one-form as $\kappa_m = G_{mn} v^n$ and unit norm means $\kappa_m v^m = 1$. Now in addition to the flat connections, we now also have contact instantons \cite{Kallen:2012cs} as saddle points, satisfying
\bea
F &=& * (\kappa \wedge F)\cr
\iota_v F &=& 0
\eea
We can add further BRST exact terms which will enhance the supersymmetry to two supercharges \cite{Bak:2015hba} and then the contact instantons can become supersymmetric solutions. Yet these instanton configurations are never localization points since the BRST exact part of the action is always nonzero on these instantons. These contributions will become exponentially suppressed and in the localization limit their contribution to the partition function becomes zero. This may sound counter-intuitive since the Yang-Mills action evaluated on contact instantons is proportional to $\frac{R}{T}$ where $R$ is the radius of $S^5/\mb{Z}_p$ and $T$ is the radius of the time-circle along which we reduce from 6d to 5d. Small instantons correspond to Kaluza-Klein modes under dimensional reduction from 6d. This suggests that the value of the classical Yang-Mills action in 5d theory could have a physical interpretation as a Kaluza-Klein momentum. But this is in contradiction with the fact that the Yang-Mills term sits in the BRST exact part of the Lagrangian and the fact that we can rescale the BRST exact part at our wish without affecting any physical observables. Then the ratio $\frac{R}{T}$ can not have any invariant significance. This is not a contradiction though, since nothing depends on radius $R$ since the theory is topological over $S^5/\mb{Z}_p$ and the ratio $\frac{R}{T}$ has no invariant physical meaning.

Assuming gauge group $SU(2)$, then on $S^5/\mb{Z}_p = L(p;q_1,q_2)$ we can have holonomies 
\bea
U &=& \begin{pmatrix}
e^{\frac{2\pi i \l}{p}} & 0\\
0 & e^{-\frac{2\pi i \l}{p}} 
\end{pmatrix}
\eea
labeled by integers subject to the $\mb{Z}_p$ identification $\l \sim \l + p$. Unlike the case with 3d CS classical action when evaluated on a flat gauge field, which gives a phase when exponentiated, here we have the 5d YM action which is vanishing on a flat gauge field. The whole action is vanishing on the flat gauge field background.

Now we can apply the localization method as follows. We write the full Lagrangian as $\L_{FCS} + c\delta V_{tot}$ where $V_{tot}=V+V'$ 
and $c>0$ is a parameter on which nothing depends. We may then take $c\rightarrow \infty$. Then the path integral localizes to localization points where $\delta V_{tot} = 0$ and its first derivative is zero. For this argument to work, we need $\delta V_{tot}\geq 0$, which, we have assured ourselves, is the case. All the sectors with nontrivial holonomies are kept since at these localization points we have $\delta V_{tot} = 0$.

It remains to compute the one-loop determinant for the fluctuations around flat gauge field backgrounds. We rescale all the fluctuation fields around the classical flat gauge field background by the factor $1/\sqrt{c}$. This has the effect of rescaling the FCS term by $1/c$. Taking $c$ large, we can neglect all higher order interaction terms in the fluctuation fields and only consider the one-loop approximation which becomes exact as we take $c$ to infinity. If we denote by $d_A = d-i[A,\cdot]$, then the fermionic operator $L$ becomes 
\bea
L &=&  i\(\begin{array}{ccc}
-*d_A/c & d_A & 0\\
- d_A^{\dag} & 0 & d_A\\
0 & - d_A^{\dag} & 0
\end{array}\)
\eea
whose square is
\bea
L^2 &=& \(\begin{array}{ccc}
d_A^{\dag}d_A/c^2 + d_A d_A^{\dag} & 0 & 0\\
0 & \triangle_1 & 0\\
0 & 0 & \triangle_0
\end{array}\)
\eea
if being understood that these Laplacians are given in terms of the background gauge field by $d_A d_A^{\dag} + d_A^{\dag} d_A$. Let us put $C=1/c^2$ that we would like take towards zero in order to localize the path integral. We now have to consider the following determinant
\bea
\det \(C\triangle_2^{coex} + \triangle_2^{ex}\) = \det \(C\triangle_2^{coex} +  \triangle_1^{coex}\) = \det \(C\triangle_2^{coex}\) \det \(\triangle_1^{coex}\)
\eea
We have the following result (see Eq.~(\ref{C}) in the appendix \ref{MP})
\bea
\det\(C\triangle_p\) = C^{-b_p} \det\triangle_p
\eea
Also, from 
\bea
\det (C\triangle_p) &=& \det (C\triangle_p^{coex}) \det (C\triangle_{p-1}^{coex})
\eea
we can iteratively deduce the scaling behavior of $\det\(C\triangle_p^{coex}\)$, starting with 
\bea
\det\(C\triangle^{coex}_0\) &=& C^{-b_0} \det\triangle^{coex}_0
\eea 
which is valid simply because $\triangle_0 = \triangle_0^{coex}$. Then we get iteratively
\bea
\det\(C\triangle^{coex}_1\) &=& C^{-b_1+b_0} \det\triangle^{coex}_1\cr
\det\(C\triangle^{coex}_2\) &=& C^{-b_2+b_1-b_0} \det\triangle^{coex}_2
\eea
and so in particular we get
\bea
\det \(C\triangle_2^{coex} + \triangle_2^{ex}\) = C^{-b_2+b_1-b_0} \det \triangle_2
\eea
We then get the following contributions to the partition function,
\bea
Z_{\phi} &=& \frac{1}{\det{}^{\frac{1}{2}}\triangle_1}\cr
Z_{YM} &=& \frac{\det{}\triangle_0}{\det{}^{\frac{1}{2}}\triangle_1}\cr
Z_{F} &=& C^{(-b_2+b_1-b_0)/4}\det{}^{\frac{1}{4}} \triangle_2 \det{}^{\frac{1}{4}} \triangle_1  \det{}^{\frac{1}{4}} \triangle_0
\eea
These are the contributions from the fields $\phi_m$, the Yang-Mills part $A_m$, and the fermionic part, respectively. We find zero modes, which we take out. Then multiplying the contributions together gives the following oscillator mode contribution to the partition function
\bea
Z_{osc} =  Z_{\phi} Z'_{YM} Z'_F = C^{(-b_2+b_1-b_0)/4} (\tau_{osc})^{\frac{1}{2}}
\eea
It now seems that $Z_{osc} \sim C^{-b_0/4} = c^{b_0/2}$ depends on a coefficient $c$ that we will take to infinity. (We will assume that $b_1=b_2=0$, and let us also recall that $b_0 = \dim H_{A}$ where $A$ represents the background gauge field and $H_A$ is the isotropy group preserving this background gauge field.) No matter how small we make $C$, the contribution from the FCS-term can not be neglected when we compute the one-loop determinant. The normal situation in supersymmetric localization is that we can neglect the contribution from the original action when we compute the one-loop determinants and only the BRST exact terms contribute to the one-loop determinants. The quadratic terms that sit in the original action are suppressed by the factor $C=1/c^2$ compared to the quadratic terms that sit in the BRST exact terms. Here the situation is different since there are no terms that are quadratic in $\psi_2$ in the BRST exact part of our Lagrangian. The leading quadratic term in $\psi_2$ is in the original FCS-term, so this contribution can not be neglected when we compute the one-loop determinant, and hence our dependence on the coefficient $c$. But in our computation we forgot to take into account a corresponding rescaling in the path integral measure. If we rescale all the modes, then this has no effect on the measure. Since our manifold is compact, the modes form a countable set. Let us label the modes by an integer $n$. Then the measure receives a product factor $\prod_{n\in \mb{Z}} \frac{1}{\sqrt{c}} = 1$ by rescaling all the modes by a factor of $\frac{1}{\sqrt{c}}$. When we take out a zero mode $n=0$, the product starts to depend on $c$ as $\prod_{n\neq 0} \frac{1}{\sqrt{c}} = \sqrt{c}$. Since there are $b_0$ zero modes in total, we get from the path integral measure the factor $c^{b_0/2}$. The precise way to see this is by expanding in mode functions that are eigenmodes of the laplace operator $\triangle_0$ on $M_5$, and then use zeta function regularization for the infinite product over nonvanishing eigenmodes (nonzero modes) of $\triangle_0$,
\bea
\prod_{n\neq 0} \frac{1}{\sqrt{c}} = \(\frac{1}{\sqrt{c}}\)^{\zeta_{\triangle_0}(0)} = c^{b_0/2}
\eea
where in the last step we used the Minakshisundaram-Pleijel theorem $\zeta_{\triangle_0}(0) = -b_0$ (see appendix \ref{MP}). Now the exponent has the wrong sign -- we need $c^{-b_0/2}$ to cancel the above $c$-dependence from the determinants. But this is actually what we have since the contribution to the zero modes comes from fermionic fields rather than from bosonic fields. For fermionic fields we have the following property $d(\frac{1}{\sqrt{c}}\psi) = \sqrt{c} d\psi$ when we rescale the field, the differential in the measure is rescaled by the inverse factor. So for fermionic fields we will encounter the product $\prod_{n\neq 0} \sqrt{c} = \frac{1}{\sqrt{c}}$ instead and this gives us the desired factor of $c^{-b_0/2}$ that cancels out all the dependence on $c$. 
  
Our Laplacian operators are defined in a flat gauge field background. Alternatively we consider fields that are satisfying twisted boundary conditions along the Hopf circle. We shall sum over all gauge inequivalent flat gauge field backgrounds. For $SU(2)$ gauge group and lens space $L(p;q_1,q_2)$ that amounts to a sum over flat gauge fields that we constructed explicitly in section 2 and which are labeled by $\l = 0,1,...,(p-1)/2$ assuming $p$ is odd. We need to multiple by a factor $Z_{zero}$ that is obtained by dividing by the volume factor \cite{Marino:2011nm}
\bea
\Vol(\H_{A^{\l}}) &=& \Vol(H_{A^{\l}}) (\Vol(M_5))^{\frac{1}{2} \dim(H_{A^{\l}})}
\eea
where $H_{A^{\l}}$ denotes the subgroup of the gauge group that leaves the background gauge field $A^{\l}$ invariant. For gauge group $SU(2)$, these volume factors are
\bea
\Vol(\G) &=& \Vol(SU(2)) (\Vol(L(p;q_1,q_2)))^{\frac{3}{2}}
\eea
for $\l=0$ and 
\bea
\Vol(\H) &=& \Vol(U(1)) (\Vol(L(p;q_1,q_2)))^{\frac{1}{2}}
\eea
for $\l>0$. The Ray-Singer torsion for $SU(2)$ gauge group is given by (\ref{SU2RS}), which we can separate into a zero mode factor and an oscillator mode factor for each $\l$ as
\bea
\tau_{0,SU(2)} &=& \[\tau_{0,zero} \tau_{0,osc}\]^3\cr
\tau_{\l,SU(2)} &=& \[\tau_{0,zero} \tau_{0,osc}\] \tau_{2\l} \tau_{-2\l}
\eea
where $\tau_{0,zero} = \frac{1}{\Vol(L(p;q_1,q_2))}$. The resulting partition function becomes precisely of the form presented in Eq (\ref{5dCS}). Unlike the case for 3d CS, for 5d FCS this one-loop result is exact by the localization principle. 

It remains the question of the normalization of the 5d FCS term, or equivalently, the question of the radius of the $SU(2)$ gauge group. We will not fully solve this problem, but will be able to derive the dependence on $k_F$, where we normalize the 5d FCS term as
\bea
\frac{ik_F}{2} \int \tr(\psi_2 \wedge d_{A} \psi_2) = \frac{ik_F}{2} (\psi_2,*d_{A} \psi_2)
\eea
This leads to a determinant
\ben
\det(ik_F *d_{A})^{1/2}\label{det}
\een
in the space of non-harmonic two-forms. The $k_F$-dependence of the determinant (\ref{det}) can be inferred from
\bea
*d_A*d_A &=& \triangle_2^{coex}
\eea
together with the formula
\bea
\det (k_F^2\triangle_2^{coex})^{1/4} &=& k_F^{(-b_2+b_1-b_0)/2} \det(\triangle_2^{coex})^{1/4}
\eea
Then for $b_2=b_1=0$ and $b_0$ being the dimension of the unbroken gauge group, this leads to the $k_F$-dependence $\sim k_F^{-b_0/2}$. Unlike the dependence on $c$ above, the dependence on $k_F$ is genuine. The reason why, is that $k_F$ multiplies the FCS term, while $c$ multiplies the BRST exact terms. Also here we did not need to rescale the field and so we did not change the path integral measure as we did when we extracted the dependence on $c$. 

Let us compare this dependence on $k_F$ with the dependence on $k$ for 3d CS perturbation theory. There we have the CS term
\bea
\frac{k}{4\pi} (B,*d_A B)
\eea
where $B$ represents a fluctuation of the gauge field around the background $A$. This leads to the determinant
\bea
\frac{1}{\det(k *d_A)} &=& \frac{1}{k^{(-b_1+b_0)/2} \det (*d_A)}
\eea
and we get the $k$-dependence $\sim k^{-b_0/2}$ (assuming that $b_1=0$), which is of the same form as we found for 5d FCS. That suggests that as an alternative to our localization compuation, we should also be able to study 5d FCS theory using perturbation theory in a small coupling $1/k_F$.

\section{Higher-dimensional knot theory}
In \cite{Guadagnini:2014mja}, abelian 3d CS was used to describe links by associating a Wilson loop with a link or a knot. Here we would like to generalize this to links and knots that are made of closed 2d surfaces embedded in a 5d manifold. The first question we should ask is what would be the definition of the Wilson surface that we should associate with such a 2d surface?

At least if the gauge group is abelian, we can dualize the 5d YM gauge potential $A_1$ into a bosonic two-form $B_2$ by taking $dB_2 = * dA_1$. We can then use this two-form to define a Wilson surface as
\bea
W(\Sigma_1) &=& \exp \(i e_i \int_{\Sigma_i} B\)
\eea
where $e_i$ is the charge associated with the loop whose trajectory forms the surface $\Sigma_i$. However, this can not be used to describe knots made of closed 2d surfaces (surfaces that form topologically nontrivial links and knot configurations) in 5d. The reason is that the linking number is anti-symmetric under exchange of two surfaces in 5d.

Instead we should use a fermionic two-form $\psi_2$. We associate an anticommuting parameter $\theta_i$ with each surface $\Sigma_i$ embedded in some five-manifold. One may think on $\theta_i$ as the fermionic analog of electric charge. The product $\theta_i \psi_2$ should have scaling dimension zero, since only then can we form a fermionic Wilson surface as
\bea
W(\Sigma_i) &=& \exp \(\theta_i \int_{\Sigma_i} \psi_2\)
\eea
This Wilson surface becomes a unitary operator without the insertion of any extra factor of $i$ in the exponent, if $\theta_i$ and $\psi_2$ are real and anticommuting. In $\mb{R}^5$ we can have two 2d closed oriented surfaces linking each other. The Gauss linking formula gives the linking number of two such surfaces $\Sigma_i$ and $\Sigma_j$ as
\bea
\lk(\Sigma_i,\Sigma_j) &\sim & \int_{\Sigma_i} dx^m \wedge dx^n \int_{\Sigma_j} dy^p \wedge dy^q \epsilon_{mnpqr} \frac{x^r-y^r}{|x-y|^5}
\eea
This linking number is anti-symmetric in 5d
\bea
\lk(\Sigma_i,\Sigma_j) &=& -\lk(\Sigma_j,\Sigma_i)
\eea
We now notice that fermionic Chern-Simons leads to the propagator
\bea
\<\psi_{mn}(x)\psi_{pq}(y)\> &\sim & \epsilon_{mnpqr} \frac{x^r-y^r}{|x-y|^5}
\eea
and so we can compute this linking number from the expectation value of two Wilson surfaces
\bea
\lk(\Sigma_i,\Sigma_j) &=& \<W(\Sigma_i) W(\Sigma_j)\>
\eea

More generally, if we have a disconnected set of surfaces $\Sigma_i$, we define our Wilson surface as 
\bea
W(\Sigma_1,...,\Sigma_n) = \exp \(\sum_{i=1}^n \theta^i \int_{\Sigma_i} \psi_2\) = \exp \(\sum_{i=1}^n \theta_i \int \psi_2 \wedge \delta_{\Sigma_i}\) = \exp\(\int \psi_2 \wedge \delta^f_{\Sigma}\)
\eea
Here we define a fermionic Poincare dual as
\bea
\delta^f_{\Sigma} &:=& \sum_{i=1}^n \theta_i \delta_{\Sigma_i}
\eea
and $\delta_{\Sigma_i}$ are the usual Poincare duals of $\Sigma_i$, defined by
\bea
\int \psi_2 \wedge \delta_{\Sigma_i} &=& \int_{\Sigma_i} \psi_2
\eea
We can now compute the expectation value of this generalized Wilson surface,
\bea
\<W(\Sigma_1,...,\Sigma_n)\> =  \int \D\psi_2 e^{\frac{i}{2} \int \(\psi_2 \wedge d\psi_2 - 2i \psi_2 \wedge \delta^f_{\Sigma}\)}
\eea
by shifting the fermionic field as
\bea
\chi &=& \psi - i \delta_{\Sigma}
\eea
We then complete the square, and we get
\bea
\<W(\Sigma_1,...,\Sigma_n)\> = \<1\> e^{\int \delta_{\Sigma}^f \wedge \delta_{D}^f} = \<1\> e^{\theta_i \theta_j \int \delta_{\Sigma_i} \wedge \delta_{D_j}} = \<1\> e^{\theta_i \theta_j \l {\rm k} (\Sigma_i,\Sigma_j)}
\eea
where we define the linking number as
\bea
\lk(\Sigma_i,\Sigma_j) &=& \int \delta_{\Sigma_i} \wedge \delta_{D_j}
\eea
with $\partial D_i = M_i$.
Now we can understand the absence of framing dependent factor in the partition function. The self-intersection is simply removed by the anticommuting property $\theta_i \theta_j = - \theta_j \theta_i$. So we do not need to consider the issue of framing to define the otherwise ambiguous self-intersection numbers.

\section{Uplift to six dimensions}
The maximal twist of the M5 brane theory amounts to identifying the R-symmetry group $SO(5)$ with the $SO(5)\subset SO(1,5)$ in the Lorentz group. Once this twist is done, we can preserve one scalar supercharge on any Lorentzian six-manifold $M_6 = \mb{R} \times M_5$ with metric 
\bea
ds^2 &=& -dt^2 + G_{mn} dx^m dx^n
\eea

The action is 
\bea
S_B &=& \frac{1}{2}\(-\frac{1}{2}(B_2, \t\triangle_2 B_2) - \frac{1}{2}(B_0,\t\triangle_0 B) + (\b{c},\t \triangle_0 c)\)\cr
&& + i (\b{b}_1,\t\triangle_1 b_1)\cr
S_{\phi} &=& \frac{1}{2} (\dot\phi,\dot\phi) - \frac{1}{2}(\phi, \triangle_1\phi)\cr
S_{\psi} &=& -\frac{i}{2}(\psi_0,\dot{\psi}_0) - \frac{i}{2}(\psi_1,\dot{\psi}_1) - \frac{i}{2}(\psi_2,\dot{\psi}_2)\cr
&& + i (\psi_1,d\psi_0) + i (\psi_2,d\psi_1)\cr
&& - \frac{i}{2} (\psi_2,*d\psi_2)
\eea
where $\t\triangle $ denotes a 6d Laplacian. Here $b_1$ and $\b{b}_1$ are ghosts for the two-form gauge field $B_2$, and $B$, $c$ and $\b{c}$ are ghosts-for-ghosts \cite{Siegel:1980jj, Kimura:1980aw}. The relative coefficients of the full action are fixed by supersymmetry and by the requirement that, upon dimensional reduction from $S^1 \times \mb{R}^5$ down to $\mb{R}^5$, we get 5d SYM theory (on flat space, twisting is trivial) with canonically normalized fields. For this twisted theory we have just one real supercharge. The term for the two bosonic ghosts $c$ and $\b{c}$ is not of the standard form. But by using the identity
\bea
\int dx dy e^{\lambda x y} = \int dx dy e^{\frac{i\lambda}{2}(x^2+y^2)} = \frac{2\pi i}{\lambda}
\eea
this term can be replaced as
\bea
(\b{c},\t\triangle_0 c) &\rightarrow& \frac{i}{2} (c,\t\triangle_0 c) + \frac{i}{2} (\b{c},\t\triangle_0 \b{c})
\eea
without changing the value of the path integral. 

The action for the fermions can be written in the form
\bea
S_{\psi} &=& -\frac{i}{2} (\Psi,\dot\Psi) + \frac{1}{2} (\Psi,L\Psi)
\eea
where $\Psi := (\psi_2,\psi_1,\psi_0)^T$ and
\bea
L &=& i\(\begin{array}{ccc}
-*d & d & 0\\
-d^{\dag} & 0 & d\\
0 & -d^{\dag} & 0
\end{array}\)
\eea
is a hermitian operator that squares to 
\bea
L^2 &=& \(\begin{array}{ccc}
\triangle_2 & 0 & 0\\
0 & \triangle_1 & 0\\
0 & 0 & \triangle_0
\end{array}\)
\eea
We will now compute the Witten index using the path integral quantization. Since there are zero modes that we take out, we need to be careful with normalization of the path integral. Our normalization of the path integral will be that which for finite dimensional integrals corresponds to  
\bea
\int \frac{dx}{\sqrt{2\pi}} e^{-\frac{\lambda}{2} x^2} &=& \frac{1}{\sqrt{\lambda}}\cr
\int d\psi d\b\psi e^{\lambda \b\psi\psi} &=& \lambda
\eea
In other words, we will get determinants without any extra multiplicative factors, when the action is canonically normalized. It turns out that all terms in our action have the canonical normalization. One can see this by computing the Dirac brackets. One then find the canonical commutation relations, which means the kinetic terms in our action are canonically normalized. This also amounts to no extra factors appear when we compute the path integral and get determinants. We get from the path integral the following results 
\bea
Z_{\psi} &=& \det\(i\partial_t + L\)^{\frac{1}{2}}\cr
&=& \det\(\partial^2_t + \triangle_2\)^{\frac{1}{4}} \det\(\partial^2_t + \triangle_1\)^{\frac{1}{4}} \det\(\partial^2_t + \triangle_0\)^{\frac{1}{4}}
\eea
\bea
Z_{\phi} &=& \frac{1}{\det\(\partial_t^2 + \triangle_1\)^{\frac{1}{2}}}
\eea
\bea
Z_{B^+} &=& \(\frac{\det\(\t\triangle_1\)}{\det\(\t\triangle_2\)^{\frac{1}{2}}
 \det\(\t\triangle_0\)^{\frac{1}{2}} \det\(-i\t\triangle_0\)}\)^{\frac{1}{2}}\cr
&=& \frac{ \det\(\partial^2_t + \triangle_1\)^{\frac{1}{4}} }{ \det\(\partial^2_t + \triangle_2\)^{\frac{1}{4}}  \det\(\partial^2_t + \triangle_0\)^{\frac{1}{4}} }
\eea
In reality the M5 brane action has a fixed coupling constant, and the canonical normalization of the action is not correct. However, when $b_2(M_6) = b_3(M_6) = 0$, those effects caused by selfduality disappear and we may assume the action has been canonically normalized by an appropriate rescaling of the fields. 

For the two-form $B$ there is a $G = U(1)$ gauge symmetry, and we have to factor out the volume $\Vol(\G)$ of the corresponding G-bundle $\G$ over the six-manifold when we compute the path integral over all gauge redundant field configurations. But after BRST gauge fixing, such a volume is factored out from the path integral, and then canceled by dividing the path integral by $\Vol(\G)$. The upshot is that we never see $\Vol(\G)$ in the final result after the cancellation of these volumes has taken place. If we multiply together all contributions, we find that all determinants cancel and we are left with 
\bea
I = Z_{B^+} Z_{\phi} Z_{\psi} = 1
\eea
We note that there are both fermionic as well as bosonic zero modes. They appear in the determinants $\det\(\partial^2_t + \triangle_0\)^{\frac{1}{4}}$ in the fermionic contribution as well as in the contribution coming from $B^+$. These zero modes are canceled since these determinants exactly cancel. Hence we do not need to remove ghost zero modes by hand and consequently we do not divide by an extra volume factor $\Vol(\G)$ as we did in the corresponding abelian 5d theory, Eq (\ref{5dvol}), where we took out ghost zero modes by hand. 

Now we have an interesting mismatch between the 6d Witten index\footnote{It is the Witten index since we assume periodic boundary conditions for the fermions around the time circle in the path integral.} $I=1$ and the corresponding 5d partition function given in eq (\ref{abelian5dFCS}). A conjecture is that 5d MSYM is precisely the same thing as 6d (2,0) theory on a circle \cite{Douglas:2010iu}, \cite{Lambert:2010iw}. Here we have only one supercharge, so we are not addressing the conjecture in its original form which keeps $16$ supercharges. But our theories are nevertheless related, so it is interesting to find a mismatch here. The origin of this mismatch lies in how a selfdual  2-form in 6d reduces to a Yang-Mills gauge field in 5d, which we can demonstrate explicitly only by assuming the gauge group is abelian.

\section{Dimensional reduction of selfdual two-form}
The 6d and 5d partition functions of a 2-form and of a 1-form potential, respectively, are\bea
Z_{6d,osc}^{(2)} &=& \frac{\det{}'\t\triangle_1}{\det{}'^{\frac{1}{2}}\t\triangle_2\det{}'^{\frac{3}{2}}\t\triangle_0}\cr
Z_{5d,osc}^{(1)} &=& \frac{\det{}'\triangle_0}{\det{}'^{\frac{1}{2}}\triangle_1}
\eea
Gauge fixing shall be extended to include any ghost zero modes as well, which then gauge fixing will take out as indicated by the primes. If we dimensionally reduce a selfdual 2-form in 6d, down to 5d, we get a 1-form gauge potential in 5d. Let us expand the $p$-form Laplace operator on the Euclidean six-manifold $M_6 = S^1 \times M_5$ as\footnote{Our sign convention is such that $\triangle = -\partial_m \partial_m$ on $\mb{R}^5$.}
\bea
\t\triangle_p
&=& -\partial_{\tau}^2+\triangle_p 
\eea
We then note the relation
\bea
\det{}' \t\triangle_p(\partial_{\tau}=0) &=& \det{}' \triangle_p \det{}' \triangle_{p-1}
\eea
This means that if we put $\partial_{\tau}=0$ to get the dimensionally reduced theory, then the 6d partition function of the selfdual 2-form reduces to 
\bea
Z^{(2+)}_{6d,osc}(\partial_{\tau}=0) &=& \frac{\det{}'^{\frac{1}{4}}\triangle_1}{\det{}'^{\frac{1}{4}}\triangle_2\det{}'^{\frac{1}{4}}\triangle_0}
\eea
where now the Laplacians are on $M_5$, and where we took the square root of the non-chiral two-form partition function. Naively we would expect to get the partition function of the 5d Maxwell theory, $Z_{5d}$. To see whether this is really true, let us form the ratio,
\bea
\frac{Z^{(2+)}_{6d,osc}(\partial_{\tau}=0)}{Z^{(1)}_{5d,osc}} &=& \frac{\det{}'^{\frac{3}{4}}\triangle_1}{\det{}'^{\frac{1}{4}}\triangle_2\det{}'^{\frac{5}{4}}\triangle_0}
\eea
This quantity is related to the oscillator mode contribution to the Ray-Singer torsion of $M_5$, which is defined as
\bea
\tau_{osc}(M_d) &=& \prod_{p=0}^d (\det{}' \triangle_p)^{-(-1)^p \frac{p}{2}}
\eea
where $d=5$ is the dimension of the manifold $M_d$ in our case. By using the relation $\det{}'\triangle_p = \det{}'\triangle_{5-p}$ which follows from the fact that $*$ commutes with the Laplacian and maps a $p$-form to a $(5-p)$-form in a one-to-one fashion, the Ray-Singer torsion becomes
\bea
\tau_{osc}(M_5) &=& \frac{\det{}'^{\frac{1}{2}}\triangle_2\det{}'^{\frac{5}{2}}\triangle_0}{\det{}'^{\frac{3}{2}}\triangle_1}
\eea
We now see that  
\ben
Z^{(2+)}_{6d,osc}(\partial_{\tau}=0) &=& \frac{1}{\sqrt{\tau_{osc}(M_5)}} Z^{(1)}_{5d,osc}\label{5dosc}
\een
This relation does not explain the emergence of the zero mode contribution to the Ray-Singer torsion. The relation that we wish to have  reads
\bea
Z^{(2+)}_{6d}(\partial_{\tau}=0) &=& \frac{1}{\sqrt{\tau_{osc}(M_5)\tau_{zero}(M_5)}} Z^{(1)}_{5d}
\eea
Then we use that $\tau_{zero}(M_5) = \frac{1}{\Vol(M_5)}$ and we get
\bea
Z^{(2+)}_{6d}(\partial_{\tau}=0) &=& \frac{1}{\sqrt{\tau_{osc}(M_5)}} \sqrt{\Vol(M_5)} Z^{(1)}_{5d}
\eea
Indeed, we have argued that the full supersymmetric partition functions are given by
\bea
Z_{5d}^{(1)} &=& \frac{1}{\sqrt{\Vol(M_5)}} Z_{5,osc}^{(1)}\cr
Z_{6d}^{(2+)} &=& Z_{6d,osc}^{(2+)}
\eea
We divide the 5d partition function by the volume of the gauge group bundle since we take out gauge zero modes. For 6d case the zero modes are canceling out so we do not divide by a correponding volume there. Now this leads us back to the relation (\ref{5dosc}).

As we show in appendix \ref{6dto5d}, this can be generalized to selfdual $2k$-form potential in $4k+2$ dimensions for $k=1,2,3,...$ where we have
\bea
Z^{(2k+)}_{4k+2,osc}(\partial_{\tau}=0) &=& \frac{1}{\sqrt{\tau_{osc}(M_{4k+1})}} Z^{(2k-1)}_{4k+1,osc}
\eea
In $4k$ dimensions ($k=1,2,3,...$) things work differently. While in $4k+2$ dimensions we have $*^2 =1$ in Lorentzian signature, in $4k$ dimensions we have $*^2=1$ in Euclidean signature. For a selfdual $2k-1$-form potential in $4k$ dimensions the relation is
\bea
Z^{(2k-1,+)}_{4k,osc}(\partial_{\tau}=0) &=& \sqrt{\tau_{osc}(M_{4k-1})} Z^{(2k-2)}_{4k-1,osc}
\eea
Let us now assume we reduce a selfdual gauge field in 4d down to a scalar field in 3d. Now the abelian scalar in 3d is not a gauge field and does not require gauge fixing, but the gauge field in 4d does require gauge fixing. So now we shall divide the volume factor on the 4d side, not on the reduced 3d side. So we have the relations
\bea
Z^{(1,+)}_{4} &=& \frac{1}{\sqrt{\Vol(M_4)}} Z^{(1,+)}_{4,osc}\cr
Z^{(0)}_{3} &=& Z^{(0)}_{3}
\eea
Since $\Vol(M_4) = R\times \Vol(M_3)$, the volume factor now combines with $\tau_{osc}$ into the full Ray-Singer torsion up to a factor of $R$, the radius of the circle along which we reduce. Note that it appears we should not not take the square root of the above volume factor as one might have expected when the gauge field is selfdual. 

Finding the nonabelian generalization of this dimensional reduction will be very interesting. We believe that the 6d Witten index is $I=1$ for any nonabelian gauge group. Knowing the 5d partition function, we may infer that the mismatch comes from reducing a nonabelian selfdual two-form to 5d nonabelian YM gauge field. This can give some clues about what is the nonabelian selfdual two-form.

\subsection*{Acknowledgements}
We would like to thank Dongmin Gang, Luca Grigoulo,  Jeong-Hyuck Park, Domenico Seminara, and Masahito Yamazaki for enlightening discussions. 
D.B. was supported in part by  IBS-R018-D2 and  NRF Grant 2017R1A2B4003095.
A.G. was supported by the grant “Geometry and Physics” from the Knut and Alice Wallenberg foundation and IBS-R018-D2.

\appendix

\section{The Ray-Singer torsion}\label{RSa}
Lecture notes on the Ray-Singer torsion are \cite{Bunke, Mnev}. Here we summarize what we need for our purposes. Let $M$ be a compact smooth oriented manifold of dimension $d$ without boundaries. 
Let $\omega_p^i$ be a basis of harmonic $p$-forms on $M$, with corresponding dual cycles $C_p^i$ in the homology of $M$. We define the inner product of two $p$-forms as
\bea
(\omega,\eta) &=& \int_M \omega \wedge * \eta
\eea
We now have a matrix $(\omega^i_p,\omega^j_p)$ for each $p$ (which may be empty if there are no harmonic $p$-forms on $M$). We define the Ray-Singer torsion as
\bea
\tau(M) &=& \tau_{zero}(M) \tau_{osc}(M)
\eea
where 
\bea
\tau_{osc}(M) &=& \prod_{p=0}^d (\det{}' \triangle_p)^{-(-1)^p \frac{p}{2}}\cr
\tau_{zero}(M) &=& \prod_{p=0}^d (v_p)^{-(-1)^p\frac{1}{2}}
\eea
and where
\bea
v_p &=& \det(\omega_p^i,\omega_p^j)
\eea

A second expression for the Ray-Singer torsion can be obtained if we decompose the Laplacian in the space of non-harmonic forms as
\bea
\triangle_p &=& \triangle_p^{coex} + \triangle_p^{ex}
\eea
Here
\bea
\triangle_p^{coex} &=& d_p^{\dag} d_p\cr
\triangle_p^{ex} &=& d_{p-1} d_{p-1}^{\dag}
\eea
where $d_p^{\dag}=\(d_p\)^{\dag}:\Omega_{p+1} \rightarrow \Omega_p$. We have
\bea
\det\triangle_p &=& \det \triangle_p^{ex} \det \triangle_p^{coex}
\eea
Since $\omega_{p-1} = d\omega_{p-2}+d^{\dag}\eta_p+\omega_{p-1}^{harm}$, we have $d\omega_{p-1} = dd^{\dag}\eta_p$ and hence only the coexact part of $\omega_{p-1}$ contributes. Therefore
\ben
\det \triangle_p^{ex} &=& \det \triangle_{p-1}^{coex}\label{extocoex}
\een
and so we have
\ben
\det \triangle_p &=& \det \triangle_p^{coex} \det \triangle_{p-1}^{coex}\label{coex}
\een
If we use (\ref{coex}) and also note that $\triangle_d^{coex}$ is trivial, we get 
\bea
\tau_{osc} &=& \prod_{p=0}^d (\det \triangle_p^{coex})^{\frac{1}{2}(-1)^p}
\eea
If $\omega_p$ is coexact, then $*\omega_p$ is exact and we have and isomorphism between coexact $p$ forms and exact $d-p$ forms that implies that
\bea
\det \triangle_p^{coex} &=& \det \triangle_{d-p}^{ex}
\eea
Moreover, by using the relation (\ref{extocoex}) we find that
\ben
\det \triangle_p^{coex} &=& \det \triangle_{d-p-1}^{coex}\label{useful}
\een
We can use this relation to get
\bea
\prod_{p=0}^d (\det \triangle_p^{coex})^{\frac{1}{2}(-1)^p} &=& \prod_{q=0}^d (\det \triangle_q^{coex})^{\frac{1}{2}(-1)^q(-1)^{d-1}}
\eea
Hence, for even dimensions $d$ we have $\tau_{osc} = 1$. Also if $d$ is even, we get $\tau_{zero} = 1$ by Poincare duality, and so $\tau = 1$. 

A third expression for the Ray-Singer torsion is expressed in terms of the Minakshisundaram-Pleijel zeta function of the Laplacian acting on $p$-forms on $M$,
\bea
\zeta_{\triangle}(s) &=& \sum_{\lambda_i \neq 0} \lambda_i^{-s}
\eea
Here the sum runs over nonzero eigenvalues $\lambda_i$ of the Laplacian. Then we define
\bea
F(s) &=& \sum_{k=0}^d - \frac{(-1)^k k}{2} \zeta_{\triangle_k}(s)
\eea
and we have  
\bea
\tau_{osc} &=& e^{-F'(0)}
\eea

We now review the proof for the metric-independence of the Ray-Singer torsion that can be found in \cite{Mnev}. The proof shows the necessary structure of $\tau_{zero}$ in the presence of zero modes. It shows that $\tau_{zero}$ has a certain ambiguity. This ambiguity can be fixed by imposing an extra condition, as we do in Eq.~(\ref{conv}). 

We assume that the dimension $d$ is odd and we begin by assuming that there are no zero modes. We note that
\bea
\zeta_{\triangle_k}(s) &=& \frac{1}{\Gamma(s)} \int_0^{\infty} dt t^{s-1} \tr_{\Omega_k}\(e^{-t\triangle_k}\)
\eea
and so we have
\bea
F(s) &=& \frac{1}{\Gamma(s)} \int_0^{\infty} dt t^{s-1} \sum_{k=0}^d -\frac{(-1)^k k}{2}\tr_{\Omega_k} \(e^{-t \triangle_k}\)
\eea
Let $u$ parametrize a one-parameter family of metrics. The Hodge duality operators depends on the metric and hence on $u$. We may emphasize that by writing it as $* =*_u$. We now define the operator
\bea
\alpha = * \dot{*} := *_u \frac{d}{du}*_u
\eea
We have 
\bea
(-1)^{k(d-k)}** &=& 1\cr
d^{\dag} &=& (-1)^{dk+d+1}*d*
\eea
which for $d$ odd yields
\bea
** &=& 1\cr
d^{\dag} &=& (-1)^k*d*
\eea
Consequently $\dot{*} * = -\alpha$ and
\bea
\dot\triangle &=& -d\alpha d^{\dag} + dd^{\dag}\alpha - \alpha d^{\dag}d + d^{\dag}\alpha d
\eea
After some computation where one uses cyclicity of trace and the fact that $d$ and $d^{\dag}$ commute with $\triangle$, it follows that 
\bea
\frac{d}{du} \sum_{k=0}^d -\frac{(-1)^k k}{2} \tr_{\Omega_k} e^{-t\triangle_k} &=& -\frac{1}{2} t \frac{d}{dt} \sum_{k=0}^d (-1)^k \tr_{\Omega_k} \(e^{-t \triangle_k} \alpha\)
\eea
and further that 
\bea
\frac{\partial}{\partial u} F(s) &=& \frac{1}{2} \sum_{k=0}^d (-1)^k  \frac{s}{\Gamma(s)} \int_0^{\infty} dt t^{s-1} \tr_{\Omega_k} \(e^{-t\triangle_k}\alpha\)
\eea
For $d$ odd we have a theorem that says that the integral of this expression has no pole in $s$ at $s=0$. Therefore we have a double zero at $s=0$ since $\Gamma(s) = 1/s + regular$. Therefore 
\bea
\frac{\partial}{\partial u} F'(0)&=& 0
\eea
proving metric independence of the analytic torsion. 

If there are zero modes, then we have to first project those out from the definition of the analytic torsion. Let $P$ be the projection from forms in $\oplus_{k=0}^d \Omega_k$ to harmonic forms in $\oplus_{k=0}^d Harm_k$. Then we replace $\tr_{\Omega_k}$ above with $\tr_{\Omega_k} (1-P)$. Formal manipulations now yield a nonzero contribution from the metric variation of the analytic torsion which is 
\bea
\frac{\partial}{\partial u} F(s) &=& -\frac{1}{2} \sum_{k=0}^d (-1)^k  \frac{s}{\Gamma(s)} \int_0^{\infty} dt t^{s-1} \tr_{Harm_k} \(\alpha\)
\eea
The integral is now elementary, and formally we have
\bea
\int_0^{\infty} dt t^{s-1} &=& -\frac{1}{s}
\eea
What we really do here is to compute the integral in the domain of $s$ where it is convergent and then we continue that result analytically in $s$. We then have
\bea
\frac{\partial}{\partial u} F(s) &=& -\frac{1}{2} \sum_{k=0}^d (-1)^k  \frac{1}{\Gamma(s)} \tr_{Harm_k} \(\alpha\)
\eea
By noting that $\Gamma(s)=1/s+reg$ we see that 
\bea
\frac{\partial}{\partial u} F'(0) &=& -\frac{1}{2} \sum_{k=0}^d (-1)^k \tr_{Harm_k} \(\alpha\)
\eea
and so now we have
\bea
\frac{\partial}{\partial u} \log \tau_{osc} = -  \frac{\partial}{\partial u} F'(0) = \frac{1}{2} \sum_{k=0}^d (-1)^k \tr_{Harm_k} \(\alpha\)
\eea
which is nonzero. We then need to add a zero mode contribution whose variation cancels the above variation to get a metric-independent Ray-Singer torsion. Let us define 
\bea
\log \tau_{zero} &=& -\frac{1}{2}\sum_{k=0}^d \sum_{i=1}^{b_k(M)} (-1)^k (\omega^k_i,\omega^k_i)_u
\eea
where $\omega^k_i$ is a metric-independent and orthonormal basis of Harmonic $k$-forms at $u=0$,
\bea
(\omega^k_i,\omega^k_j)_{u=0} &=& \delta_{ij}
\eea
where we define
\bea
(\omega,\omega)_u &=& \int_M \omega \wedge *_u \omega
\eea
Then
\bea
\frac{\partial}{\partial u} (\omega,\omega)_u &=& \int_M \omega \wedge \dot{*} \omega \cr
&=& \int_M \omega \wedge **\dot{*}\omega\cr
&=& \int_M \omega \wedge *\alpha \omega
\eea
and so we find its metric variation evaluated at $u=0$ as
\bea
\frac{\partial}{\partial u} \log \tau_{zero} = -\frac{1}{2}\sum_{k=0}^d \sum_{i=1}^{b_k(M)} (-1)^k \tr_{Harm_k} \(\alpha\)
\eea
which precisely cancels the metric variation of the analytic torsion. This completes the proof.

However, $\tau_{zero}$ is not uniquely fixed by the requirement that $\omega^k_i$ form an orthonormal basis at $u=0$, since there are many ways that we can introduce a parameter $u$ and the metric is not uniquely fixed by the condition $u=0$. To improve this situation, we will fix the point $u=0$ by the requirement
\ben
\int_M *_{u=0} 1 &=& 1\label{conv}
\een

\subsection{Explicit computations}\label{RSE}
Let us compute the Ray-Singer torsion for a circle with the metric $ds^2 = r^2 d\theta^2$. The Hodge operator acts as $*1 = r d\theta$ and $*d\theta = \frac{1}{r}$. We have
\bea
\tau_{osc} &=& \sqrt{\prod_{n\neq 0} \(\frac{n}{r}\)^2}
\eea
which can be computed using zeta function regularization with the result
\bea
\tau_{osc} &=& 2\pi r
\eea
Note that $\tau_{osc}$ depends on the metric. We need to multiply by $\tau_{zero}$ to get a metric-independent result. On $S^1$ there are 0-form and 1-form zero modes
\bea
\omega_0 &=& 1\cr
\omega_1 &=& \frac{d\theta}{2\pi}
\eea
These are chosen such that they are orthonormal at $r_0=\frac{1}{2\pi}$ where the circumference is one, $2\pi r_0 = 1$, and hence corresponds to the point $u=0$. For example, we could let $r = \frac{1}{2\pi} + u$. For a generic radius $r$, we get 
\bea
(\omega_0,\omega_0) &=& 2\pi r\cr
(\omega_1,\omega_1) &=& \frac{1}{2\pi r}
\eea
Then 
\bea
\tau_{zero} &=& \frac{1}{2\pi r}
\eea
and we get
\bea
\tau = \tau_{zero} \tau_{osc} = 1
\eea
which is independent of the metric. 

Let us now compute the Ray-Singer torsion for $S^5$. First we compute
\bea
\tau_{osc} &=& \frac{\det'\triangle_0^{coex} \det'^{\frac{1}{2}}\triangle_2^{coex}}{\det'\triangle_1^{coex}}
\eea
We can compute this knowing the eigenvalues,
\bea
(\lambda_0)_n &=& \frac{1}{r^2}n(n+4)\cr
(\lambda_1)_n &=& \frac{1}{r^2}(n+1)(n+3)\cr
(\lambda_2^+)_n &=& \frac{1}{r^2}(n+2)^2
\eea
and the degeneracies
\bea
(b_0)_n &=& \frac{1}{12}(n+1)(n+2)^2(n+3)\cr
(b_1)_n &=& \frac{1}{3}n(n+2)^2(n+4)\cr
(b_2^+)_n &=& \frac{1}{4}n(n+1)(n+3)(n+4)
\eea
of the spherical harmonics. Here $n = 0,1,2,...$ and we see that we have one zero eigenvalue $(\lambda_0)_0=0$ with degeneracy $(b_0)_0=1$. We will take out this zero mode. Furthermore, since $(b_1)_0=0$ and $(b_2^+)_0=0$, the contribution from the remaining modes with $n=0$ gives simply a multiplicative factor of $1$ that we can forget about. Then the rest becomes 
\bea
\ln \tau_{osc} &=& \sum_{n=1}^{\infty} \[(b_0)_n \(\ln \frac{n}{r} + \ln\frac{n+4}{r}\) + (b^+_2)_n 2 \ln\frac{n+2}{r} - (b_1)_n \(\ln\frac{n+3}{r}+\ln\frac{n+1}{r}\)\]
\eea
Before we apply zeta function regularization, we shift $n$ such that we get the sum in the form
\bea
\ln \tau_{osc} &=& \sum_{n=1}^{\infty} \((b_0)_{n+4}+(b_0)_n+2(b_2^+)_{n+2}-(b_1)_{n+1}-(b_1)_{n+3}\)\ln\frac{n+4}{r}\cr
&&+ (b_0)_1 \ln \frac{1}{r} + (b_0)_2 \ln \frac{2}{r} + (b_0)_3 \ln \frac{3}{r} + (b_0)_4 \ln \frac{4}{r}\cr
&& + (b_2)_1 2 \ln \frac{3}{r} + (b_2)_2 2\ln \frac{4}{r}\cr
&& - (b_1)_1 \ln \frac{4}{r} - (b_1)_1 \ln \frac{2}{r} - (b_1)_2 \ln \frac{3}{r} - (b_1)_3 \ln \frac{4}{r}
\eea
Using Mathematica we find that this simplifies to 
\bea
\ln \tau_{osc} &=& \sum_{n=1}^{\infty} 6 \ln \frac{n+4}{r}\cr
&& + 5 \ln \frac{2}{r} + 6 \ln \frac{3}{r} +5 \ln \frac{4}{r}\cr
&=& 6\(\sum_{n=1}^{\infty} \ln \frac{n}{r}\) - \ln \frac{8}{r^2}
\eea
Now we apply zeta function regularization on the infinite sum, and get
\bea
\tau_{osc} &=& \pi^3 r^5
\eea
We notice that this is the volume of $S^5$. We shall now choose normalization for our harmonic forms on $S^5$. These are
\bea
\omega_0 &=& 1\cr
\omega_5 &=& \frac{\h\Omega_5}{\pi^3}
\eea
where $\h\Omega_5$ denotes the volume-form of the unit five-sphere. Then 
\bea
(\omega_0,\omega_0) &=& \pi^3 r^5\cr
(\omega_5,\omega_5) &=& \frac{1}{\pi^3 r^5}
\eea
and we get
\bea
\tau_{zero} = (\omega_0,\omega_0)^{-\frac{1}{2}} (\omega_5,\omega_5)^{\frac{1}{2}} = \frac{1}{\pi^3 r^5}
\eea
and so we get
\bea
\tau &=& 1
\eea
For $S^1/\mb{Z}_p$ which is again a circle, we have shown that we have $\tau=1$. 

Let us now proceed to $S^5/\mb{Z}_p=L(p;1,1)$. For the zero mode part, as our harmonic $p$-forms, we take 
\bea
\omega_0 &=& 1\cr
\omega_5 &=& \frac{p \h\Omega_5}{\pi^3}
\eea
Then 
\bea
(\omega_0,\omega_0) &=& \frac{\pi^3 r^5}{p}\cr
(\omega_5,\omega_5) &=& \frac{p}{\pi^3 r^5}
\eea
From this we get  
\bea
\tau_{zero} &=& \frac{p}{\pi^3 r^5}
\eea
Let us now turn to the oscillator modes (and let us temporarily put the radius $r=1$). We define
\bea
d(p,q) &=& \frac{1}{2} (p+1)(q+1)(p+q+2)
\eea
as the dimension of the representation labeled $(p,q)$ of $SU(3)$. We then introduce the following refined dimensions of representations  of $SU(4)=SO(6)$ \cite{Bak:2016vpi}, 
\bea
d(n,a,0) &=& \sum_{k=0}^n d(k,n-k) e^{i a (2k-n)}\cr
d(n,a,1) &=& \sum_{k=0}^{n-1} \Big[(d(k,n-k-1) e^{i a(2k-n+1)} + d(k-1,n-k-1) e^{i a(2k-n-1)} \cr
&&+ d(k+1,n-k) e^{i a(2k-n+1)} + d(k,n-k) e^{i a (2k-n+3)}\Big]\cr
d(n,a,2) &=& \sum_{k=0}^{n-1} \Big[d(k,n-k-1) e^{i a(2k-n-2)} + d(k,n-k) e^{i a(2k-n)} + d(k,n-k+1) e^{i a(2k-n+2)}\Big]
\eea
We then define
\bea
d(n,p,rank) &=& \frac{1}{p} \sum_{\l=0}^{p-1} d\(n,\frac{2\pi \l}{p},rank\)
\eea
For $p=1$ these are 
\bea
d(n,1,0) &=& (b_0)_n\cr
d(n,1,1) &=& (b_1)_n\cr
d(n,1,2) &=& (b_2^+)_n
\eea
and for $p=2$ they are 
\bea
d(n,2,0) &=& \frac{1}{2}\(1+(-1)^n\)(b_0)_n\cr
d(n,2,1) &=& \frac{1}{2}\(1-(-1)^n\)(b_1)_n\cr
d(n,2,2) &=& \frac{1}{2}\(1+(-1)^n\)(b_2^{+})_n
\eea
and for higher values of $p$ we may obtain corresponding, but much more complicated, expressions for the dimensions of the representations for spherical harmonics on $S^5/\mb{Z}_p$. We notice that the result for $p=2$ reflects the fact that we keep those spherical harmonics which are even under $z_i \rightarrow -z_i$ if we embed $S^5$ into $\mb{C}^3$ with complex coordinates $z_i$ ($i=1,2,3$). For the scalar and two-form, these are spherical harmonics of even degree, while for the vector spherical harmonics, those are of odd degree $n$.

We then define
\bea
D(n,p) &=& d(n+4,p,0) + d(n,p,0) + 2 d(n+2,p,2) - d(n+1,p,1) - d(n+3,p,1)
\eea
and
\bea
D(p) &=& d(2,p,0)\ln(2)+d(3,p,0)\ln(3)+d(4,p,0)\ln(4)\cr
&&+2 d(1,p,2)\ln(3) + 2 d(2,p,2) \ln(4)\cr
&&-d(1,p,1)\ln(4) - d(1,p,1)\ln(2) - d(2,p,1)\ln(3) - d(3,p,1)\ln(4)
\eea

We computed these quantities up to $p=4$ with Mathematica. If we define $\omega_p := \exp\frac{2\pi}{p}$, then we can express the results as
\bea
D(n,1) &=& 6\cr
D(n,2) &=& 6\frac{1}{2} \(1+\omega_2^n\)\cr
D(n,3) &=& 6\frac{1}{3} \(1+\omega_3^{n-2}+\omega_3^{2(n-2)}\)\cr
D(n,4) &=& 6\frac{1}{4}\(1+\omega_4^n+\omega_4^{2n}+\omega_4^{3n}\)
\eea
and 
\bea
D(1) &=& 5 \ln(2) + 6 \ln(3) + 5 \ln(4)\cr
D(2) &=& 5 \ln(2) + 5 \ln(4)\cr
D(3) &=& -\ln(2) + 6 \ln(3) - \ln(4)\cr
D(4) &=& -\ln(2) + 5 \ln(4)
\eea
Also, for $n\geq 4$, we get
\bea
D(n) &=& -\ln(2)-\ln(4)
\eea
Putting these results together, we find that, at least up to $p=4$,
\bea
\ln \tau_{osc} &=& -\ln(8)+6\sum_{k=1}^{\infty} \ln(pk)
\eea
which by zeta function regularization leads to
\bea
\tau_{osc}  &=& \frac{\pi^3 r^5}{p^3}
\eea
We believe this formula is valid for any positive integers $p$ although we have checked it only for $p=1,2,3,4$. Combining this with the zero mode contribution $\tau_{zero}$, we get
\bea
\tau &=& \frac{1}{p^2}
\eea
Our results on $S^1/\mb{Z}_p$ and $S^5/\mb{Z}_p$ are now consistent with a general formula for the Ray-Singer torsion on $S^{2N-1}/\mb{Z}_p$,
\bea
\tau &=& \frac{1}{p^{N-1}}
\eea
which we stated as a conjecture in the main text as Eq.~(\ref{general}).

\section{The Minakshisundaram-Pleijel theorem}\label{MP}
The Minakshisundaram-Pleijel theorem \cite{b} says that when $d$ is odd, the number of zero modes of the Laplacian acting on the space of $p$-forms, is encoded in the spectrum of the non-harmonic forms,
\ben
\zeta_{\triangle_p}(0) &=& -b_p\label{MP}
\een
where $b_p = \dim\Ker\triangle_p$. The regularized value for the determinant of $\triangle_p$ is given by 
\bea
\det\(\triangle_p\) &=& e^{-\zeta_{\triangle_p}'(0)}
\eea
By noting that for $C \in \mb{C}$
\bea
\zeta_{C \triangle_p}'(s) &=& \(\zeta_{\triangle_p}(s)\log C + \zeta_{\triangle_p}'(s)\) C^{-s}
\eea
we get
\ben
\det\(C \triangle_p\) = C^{\zeta_{\triangle_p}(0)} \det\(\triangle_p\) = C^{-b_p} \det\(\triangle_p\)\label{C}
\een

\section{Partial gauge fixing by the Faddeev-Popov method}\label{gf1}
Here we illustrate a general theorem in \cite{Adams:1996hi} by a very simple example that we borrow from the appendix in \cite{Rozansky:1993zx}. We consider the `path integral' in zero dimensions with target space $\mb{R}^2$,
\bea
Z &=&  \frac{1}{\Vol(G)} \int_{\mb{R}^2} \frac{dX}{\sqrt{2\pi}} \frac{dY}{\sqrt{2\pi}} e^{-K S(R)}
\eea
where $G=U(1)$ is a gauge symmetry and $\Vol(G)=2\pi$ is its volume. If we assume that $R_0>0$ is a minimum for the `action' $S(R)$, then the saddle point approximation gives 
\bea
Z &=& \frac{R_0 e^{-K S(R_0)}}{\sqrt{2\pi K S''(R_0)}}
\eea
which is a good approximation when $K$ is large. 

On the other hand, we can use the $U(1)$ gauge symmetry that acts on the `fields' as
\bea
\begin{pmatrix} X^{\Lambda}\\ Y^{\Lambda} \end{pmatrix} &=& \begin{pmatrix} \cos \Lambda & \sin\Lambda \\ -\sin\Lambda & \cos\Lambda \end{pmatrix} 
\begin{pmatrix} X \\Y \end{pmatrix}
\eea
to fix the gauge $X = 0$. By the Faddeev-Popov procedure, we begin by defining a gauge fixing function
\bea
G^{\Lambda} &=& K R_0 X^{\Lambda}
\eea
From
\bea
1 = \int dG^{\Lambda} \delta(G^{\Lambda}) = \int d\Lambda \frac{dG^{\Lambda}}{d\Lambda} \delta(G^{\Lambda}) = K R_0 \int d\Lambda Y^{\Lambda} \delta(G^{\Lambda}) = K R_0^2 \int d\Lambda \delta(G^{\Lambda})
\eea
we read off the FP determinant
\bea
\det\triangle_{FP} &=& K R_0^2
\eea
Inserting $1$ into the path integral, we get
\bea
Z = K R_0^2 \int \frac{dX}{\sqrt{2\pi}} \frac{dY}{\sqrt{2\pi}} e^{-K S(R)} \delta(K R_0 X)
\eea
We write 
\bea
\delta(K R_0 X) = \int dZ e^{2\pi i K R_0 Z X}
\eea
and expand about $R_0$ to get
\bea
Z = K R_0^2 e^{-K S(R_0)} \int \frac{dX}{\sqrt{2\pi}} \frac{dY}{\sqrt{2\pi}} dZ e^{-\frac{1}{2} K S''(R_0) (X^2+Y^2)} e^{2\pi i K R_0 Z X}
\eea
The action involves the following matrix
\bea
L &=& K \begin{pmatrix} S''(R_0) & 0 & 0\\
0 & S''(R_0) & 2\pi i R_0\\
0 & 2\pi i R_0 & 0
\end{pmatrix}
\eea
Then 
\bea
Z = \frac{\det\triangle_{FP}}{\sqrt{\det(L)}} e^{-K S(R_0)} = \frac{R_0 e^{-K S(R_0)}}{\sqrt{2\pi K S''(R_0)}} 
\eea
which agrees with the saddle point approximation. 

Let us now instead assume that the global minimum of the action as at $R_0=0$. In this case 
\bea
\det\triangle_{FP} &=& 0
\eea
and 
\bea
L &=& K \begin{pmatrix} S''(R_0) & 0 & 0\\
0 & S''(R_0) & 0\\
0 & 0 & 0
\end{pmatrix}
\eea
and the general formula
\bea
Z = \frac{\det\triangle_{FP}}{\sqrt{\det(L)}} e^{-K S(R_0)} 
\eea
becomes ill-defined since there is a fermionic ghost zero mode of $\triangle_{FP}$ as well as a bosonic zero mode in $L$. As a first try, we take out all those zero modes. Then we get
\bea
\det{}'\triangle_{FP} &=& 1
\eea
(we shall define the determinant of an empty FP matrix to be $1$ since that means we are not gauge fixing anything) and 
\bea
L' &=& K \begin{pmatrix} S''(R_0) & 0\\
0 & S''(R_0)\\
\end{pmatrix}
\eea
and we would arrive at the result
\bea
Z = \frac{\det{}'\triangle_{FP}}{\sqrt{\det{}'(L)}} e^{-K S(R_0)} 
\eea
where primes indicate that the zero modes are taken out. It turns out that this gives almost the correct answer. We can compute $Z$ for large $K$ without gauge fixing the $U(1)$ gauge symmetry at all,
\bea
Z &=& \frac{1}{\Vol(G)} e^{-K S(0)} \int \frac{dX}{\sqrt{2\pi}} \frac{dY}{\sqrt{2\pi}} e^{-\frac{K}{2} S''(0) \(X^2+Y^2\)} 
\eea
In this case the FP determinant is trivial,
\bea
\det \triangle_{FP} &=& 1
\eea
while the $L$ matrix is given by
\bea
L &=& K \begin{pmatrix} S''(0) & 0\\
0 & S''(0)
\end{pmatrix}
\eea
Then the result can be expressed as
\bea
Z &=& \frac{1}{\Vol(G)} \frac{\det\triangle_{FP}}{\sqrt{\det(L)}} e^{-K S(0)}
\eea
Thus what we were missing above when we took out the zero modes, was to divide by a volume factor $\Vol(G)$.

This simple 2d example shows two special cases of a more general result \cite{Adams:1996hi}. When the gauge symmetry is fully gauge fixed by a saddle point solution (in the above example, $X=R_0>0$), we get 
\bea
Z &=& e^{-S(\mbox{saddle point})} \times ({\mbox{one-loop determinants}})
\eea
If on the other hand the gauge symmetry is not gauge fixed at all by the saddle-point solution (in the above example, $X=Y=0$), we get
\bea
Z &=& \frac{1}{\Vol(\G)} \times e^{-S(\mbox{saddle point})} \times ({\mbox{one-loop determinants}})
\eea
In quantum field theories, the gauge symmetry is an infinite-dimensional local symmetry at each point on the manifold and then we denote such a gauge symmetry as $\G$ which is a $G$-bundle over the manifold. But we can consider quantum theories that are not field theories, and whose gauge symmetry is not a local symmetry but can be any reduntant description of the quantum theory. For gauge groups bigger than $SO(2)$, there can also be intermediate cases where the gauge symmetry is only partially gauge fixed. For those cases we have
\bea
Z &=& \frac{1}{\Vol(\H)} \times e^{-S(\mbox{saddle point})} \times ({\mbox{one-loop determinants}})
\eea
where $\H$ is the subgroup of the gauge symmetry $\G$ that is not gauge fixed by the saddle point solution. To illustrate such a case we need a bigger gauge group than $U(1)$ in order to have a proper subgroup. Let us consider an example with $SO(3)$ gauge symmetry,
\bea
Z &=& \frac{1}{\Vol(G)} \int \frac{dX}{\sqrt{2\pi}}\frac{dY}{\sqrt{2\pi}}\frac{dZ}{\sqrt{2\pi}} e^{-K S(R)}
\eea
The saddle point approximation gives
\bea
Z &=& \frac{1}{\Vol(G)} \frac{2 R_0^2 e^{-KS(R_0)}}{\sqrt{KS''(R_0)}}
\eea
The gauge group $G=SO(3)$ that acts as
\bea
\begin{pmatrix} X^{\Lambda}\\ Y^{\Lambda}\\ Z^{\Lambda} \end{pmatrix} &=& \begin{pmatrix} \cos\alpha & -\sin\alpha & 0\\
\sin\alpha & \cos\alpha & 0\\
0 & 0 & 1\end{pmatrix} \begin{pmatrix} \cos\beta & 0 & \sin\beta\\
0 & 1  & 0 \\
-\sin\beta & 0 & \cos\beta\end{pmatrix} \begin{pmatrix} \cos\gamma & -\sin\gamma & 0\\
\sin\gamma & \cos\gamma & 0\\
0 & 0 & 1
\end{pmatrix} 
\begin{pmatrix} X\\ Y\\ Z \end{pmatrix}
\eea
Here we have the $SO(3)$ coordinate ranges, $\alpha\in [0,2\pi]$, $\beta\in [0,\pi]$ and $\gamma\in [0,2\pi]$. If we fix the gauge $X=Y=0$, there will be a residual gauge symmetry $H = SO(2)$ whose rotations are parametrized by the angle $\gamma$. The rotations of the points at $X=Y=0$ (the north and the south poles) are given by
\bea
\begin{pmatrix} X^{\Lambda}\\ Y^{\Lambda}\\ Z^{\Lambda} \end{pmatrix} &=& \begin{pmatrix} Z \cos\alpha \sin\beta  \\
Z \sin\alpha\sin\beta \\
Z \cos\beta
\end{pmatrix}  
\eea
To fix the gauge partially by imposing $X=Y=0$, we define two gauge fixing functions 
\bea
G^{\Lambda}_1 &=& K R_0 X^{\Lambda}\cr
G^{\Lambda}_2 &=& K R_0 Y^{\Lambda}
\eea
We have
\bea
1 = \int dG^{\Lambda}_1 dG^{\Lambda}_2 \delta(G^{\Lambda}_1) \delta(G^{\Lambda}_2) = \int d\alpha d\beta J  \delta(G^{\Lambda}_1) \delta(G^{\Lambda}_2) 
\eea
where $J$ is the Jacobian
\bea
J &=& |K^2 R_0^2 \(\partial_{\alpha} X^{\Lambda} \partial_{\beta} Y^{\Lambda} - \partial_{\beta} X^{\Lambda} \partial_{\alpha} Y^{\Lambda}\)|
\eea
At the points $X=Y=0$ this becomes
\bea
J &=& K^2 R_0^4 |\sin\beta\cos\beta|
\eea
We then write
\bea
d\alpha d\beta d\gamma J = K^2 R_0^4 \D\Lambda \cos\beta
\eea
where 
\bea
\D\Lambda = d\alpha d\beta d\gamma \sin\beta
\eea
is the Haar measure of $SO(3)$. Then we make a gauge rotation of the action and use the gauge invariance, which enables us to isolate an integral over the Haar measure alone, and put $\alpha=\beta=\gamma=0$ everywhere else. This way we get
\bea
\det\triangle_{FP} &=& K^2 R_0^2
\eea
and then we end up with the result
\bea
Z = \frac{1}{\Vol(H)} \frac{\det\triangle_{FP}}{\sqrt{\det(L)}} e^{-K S(R_0)}
\eea
where
\bea
L &=& K\begin{pmatrix} 0 & 0 & 2\pi i R_0 & 0 & 0\\
0 & 0 & 0 & 2\pi i R_0 & 0 \\
2\pi i R_0 & 0 & 0 & 0 & 0 \\
0 & 2\pi i R_0 & 0 & 0 & 0 \\
0 & 0 & 0 & 0 & S''(R_0)
\end{pmatrix}
\eea
and 
\bea
\Vol(H) = \int_0^{2\pi} d\gamma
\eea
By explicitly computing this expression for $Z$, we reproduce the result of the saddle-point approximation.

\section{Gauge fixing of zero modes}\label{gf}
Gauge fixing of fermionic and bosonic zero modes has been analysed in \cite{Blau:1989bq}. This method has reappeared more recently in supersymmetric localization \cite{Pestun:2007rz, Kallen:2011ny}. Our topological field theories in 6d and 5d consist of fields with corresponding ghost hierarchy that are all $p$-forms of various degrees, either fermionic or bosonic. Let us assume the gauge group is abelian. Then by Hodge decomposition, any bosonic $p$-form can be decomposed into a coexact, an exact and a harmonic piece,
\bea
A_p &=& d^{\dag} \alpha_{p-1} + d \beta_{p-1} + \gamma_p
\eea

\subsection{Bosonic zero mode gauge fixing}
If the action has a gauge symmetry $\delta A_p = d\Lambda_{p-1}$, then $\beta_{p-1}$ is projected out by gauge fixing. There can also be zero modes, which we will treat in a similar way as the above gauge symmetries. A zero mode for $A_p$, means that the action is invariant under $\delta A_p = \Lambda_p$ where $\Lambda_p$ is harmonic. We treat this as a gauge symmetry that we gauge fix by adding the Lagrange multiplier term $\epsilon_i (\omega^i_p,A)$ to the action. Here $\epsilon_i$ are bosonic constant Lagrange multipliers, $\omega^i_p$ is some metric-independent choice of basis for the space of harmonic $p$-forms. Integrating over $\epsilon_i$ imposes the delta function constraint $(\omega^i_p,A) = 0$, which means the harmonic piece $\gamma_p$ is projected out in a BRST invariant manner. Here the BRST variations are 
\bea
\delta A_p &=& \omega^i_p c_i\cr
\delta c_i &=& 0\cr
\delta \bar{c}_i &=& \epsilon_i\cr
\delta \epsilon_i &=& 0
\eea
where $\delta$ changes the Grassmann properties of the fields, $c_i$, $\bar{c}_i$ and $\epsilon_i$ are all constants. The full BRST exact gauge fixing term is 
\bea
\delta (\bar{c}_i \omega^i_p,A_p) = \epsilon_i (\omega^i_p,A_p) - \bar{c}_i c_j (\omega^i_p,\omega^j_p)  
\eea
We then first consider the path integral over the bosonic zero modes
\bea
\int [\N_B dA_i] [\N_B d\epsilon_i] e^{\epsilon_i A_j (\omega_p^i,\omega_p^j)} &=& \N_B^2 \int [dA_i] 2\pi \delta(A_j (\omega_p^i,\omega_p^j))\cr
&=& \frac{2\pi}{v_p} \N_B^2\int [dA^i] \delta(A^i)\cr
&=& \frac{2\pi}{v_p} \N_B^2
\eea
where $v_p = \det(\omega_p^i,\omega_p^j)$ is the Jacobian that is produced as we change variables from $A_i$ to $A^i = (\omega^i_p,\omega^j_p)A_j$ in the measure. Next we consider the path integral over the fermionic zero modes
\bea
\int [\N_F dc_i] [\N_F d\b{c}_i] e^{\bar{c}_i c_j (\omega^i_p,\omega^j_p)} &=& v_p \N_F^2
\eea
Multiplying together, we get
\bea
2\pi (\N_B \N_F)^2
\eea

\subsection{Fermionic zero mode gauge fixing}
If instead the $p$-form is a fermionic field $\psi_p$ with the symmetry $\delta \psi_p = \lambda_p$ where $\lambda_p$ is a fermionic harmonic $p$-form, then we add the Lagrange multiplier term $\epsilon_i (\omega^i_p,\psi_p)$ to the action where now $\epsilon_i$ are fermionic constant parameters. BRST variations are
\bea
\delta \psi_p &=& a_i \omega_p^i\cr
\delta a_i &=& 0\cr
\delta \bar{a}_i &=& \epsilon_i\cr
\delta \epsilon_i &=& 0
\eea
where $\epsilon_i$ are fermionic zero modes, $a_i, \b{a}_i$ are bosonic zero modes. We add the BRST-exact term
\bea
\delta \big(\N_1(\b{a}_i\omega_p^i,\psi) + \N_2(a_i\omega_p^i,\psi)\big) &=& \N_1\epsilon_i (\omega_p^i,\psi) + \N_1\b{a}_i a_j (\omega_p^i,\omega_p^j) + \N_2 a_i a_j (\omega_p^i,\omega_p^j)\cr
&=& \big(\N_1\epsilon_i \psi_j + \N_1\b{a}_i a_j + \N_2 a_i a_j\big) (\omega^i_p,\omega^j_p)
\eea
The path integral over the fermionic zero modes is
\bea
Z_0 &=& \int [\N_F d\psi_i] [\N_F d\epsilon_i] \exp\[-\big(\N_1\epsilon_i \psi_j + \N_1\b{a}_i a_j + \N_2 a_i a_j\big) (\omega^i_p,\omega^j_p)\]\cr
&=& (\N_F)^2 \N_1 v_p \exp\[-\big(\N_1\b{a}_i a_j + \N_2 a_i a_j\big) (\omega^i_p,\omega^j_p)\]
\eea
We complete the square,
\bea
(\omega^i_p,\omega^j_p) \(\N_1\b{a}_i a_j + \N_2 a_i a_j\) &=& \N_2 (\omega^i_p,\omega^j_p) \(a_i + \frac{\N_1}{2\N_2} \b{a}_i\) \(a_j + \frac{\N_1}{2\N_2} \b{a}_j\) - \frac{(\N_1)^2}{4\N_2} (\omega^i_p,\omega^j_p) \b{a}_i \b{a}_j
\eea
The Gaussian integral is convergent for $\N_2>0$ and $\N_1$ purely imaginary. At such values we can compute the Gaussian integrals over the bosonic zero modes 
\bea
\int [\N_B da_i] [\N_B d\b{a}_i] \exp \(-\N_2 \(a_i + \frac{\N_1}{2\N_2} \b{a}_i\)^2 + \frac{(\N_1)^2}{4\N_2} (\b{a}_i)^2\)
\eea
and get the result
\bea
Z_0 = (\N_B \N_F)^2 \N_1 v_p \sqrt{\frac{\pi}{\N_2 v_p}} \sqrt{\frac{\pi}{\frac{(\N_1)^2}{4\N_2}v_p}} = 2\pi (\N_B\N_F)^2
\eea
We see that all the dependence on $\N_1$ and $\N_2$ cancels out. This was known by general considerations since the added term was BRST-exact, but it is nevertheless nice to see how this happens by an explicit computation. At other values of $\N_1$ and $\N_2$ we define the path integral by analytic continuation. Since it is just a constant, the analytic continuation of the path integral is trivial -- it will remain to be equal to this constant value for all values on $\N_1$ and $\N_2$. 

\subsection{Fermionic zero mode gauge fixing, once again}
As was noted in \cite{Blau:1989bq}, this method does not work for all the $p$-forms in a ghost hierarchy. To quote \cite{Blau:1989bq}: `This works well for the ghosts that are present on the right-hand ledge of the ghost-triangle.' To illustrate what is meant by this, let us consider as an example Maxwell theory with the nonharmonic BRST variations
\ben
\delta A &=& d c\cr
\delta \bar{c} &=& iB\cr
\delta B &=& 0\cr
\delta c &=& 0\label{MW}
\een
There are two ghosts $c$ and $\bar{c}$, but only the ghost $c$ is on the right-ledge of the ghost-triangle. Let us assume these have zero-form harmonics with corresponding BRST variations. For the $c$ ghost, these will be 
\bea
\delta c &=& a_i \omega_0^i\cr
\delta a_i &=& 0\cr
\delta \bar{a}_i &=& \epsilon_i\cr
\delta \epsilon_i &=& 0
\eea
which remain nilpotent also when combined with (\ref{MW}). But for the $\bar{c}$ ghost we already have BRST transformations from the above
\bea
\delta \bar{c} &=& i B\cr
\delta B &=& 0
\eea
which can be extended to include harmonic parts as well. We then enlarge this by adding (constant) ghosts $\bar{\sigma}$ and $\tau$ whose BRST variations are
\bea
\delta \bar{\sigma}_i &=& i \tau_i\cr
\delta \tau_i &=& 0
\eea
Then we add the BRST exact term 
\bea
\delta (\bar\sigma_i\omega^i_0, \bar{c}) = i(\tau_i\omega^i_0,\bar{c}) + i(\bar\sigma_i\omega^i_0,B)
\eea
When we integrate over the fermionic zero modes, we get
\bea
\int [\N_F d\tau_i] [\N_F d\b{c}_i] e^{i\tau_i\bar{c}_j(\omega_0^i,\omega_0^j)} &=& i v_p (\N_F)^2
\eea
For the bosons, we get
\bea
\int [\N_B d\sigma_i] [\N_B dB_i] e^{i\b{\sigma}_i B_j (\omega_0^i,\omega_0^j)} &=& (\N_B)^2 \int [dB_i] 2\pi\delta(B_j (\omega_0^i,\omega_0^j))\cr
&=& \frac{2\pi (\N_B)^2}{v_p} 
\eea
Multiplying together, we get
\bea
2 \pi (\N_F\N_B)^2 
\eea

If we choose the path integral measure for the zero modes such that $\N_F \N_B = \frac{1}{\sqrt{2\pi}}$, then we can summarize our result as follows: removing any set of harmonic $p$-form zero modes from the path integral in a BRST invariant way, always produces the same factor $2\pi (\N_F \N_B)^2 = 1$ no matter the zero mode is bosonic or fermionic, or on the right-ledge of the ghost-triangle or not. All sets of harmonic zero modes produce the same factor.

\section{A review of 3d Chern-Simons perturbation theory}\label{3dCS}
Here we review what we will need from \cite{Witten:1988hf, Beasley:2005vf, Rozansky:1993zx}. The starting point is the Chern-Simons action
\bea
S(A) &=& \frac{k}{4\pi} \int \tr\(A \wedge dA - \frac{2i}{3} A^3\)
\eea
If we define the covariant derivative as
\bea
D_m^A &=& \nabla_m - i [A_m,\cdot]
\eea
then a gauge transformation associated with the group element $g$ will act as 
\bea
A_m &\rightarrow & A_m^g\cr
A_m^g &=& i g^{-1} \nabla_m g + g^{-1} A_m g
\eea
This can also be expressed as
\bea
D_m^{A^g} &=& g^{-1} D_m^A g
\eea
We have BRST variations 
\bea
\delta A_m &=& D_m c\cr
\delta c &=& \frac{i}{2} \{c,c\}\cr
\delta B &=& 0\cr
\delta \b{c} &=& i B
\eea
The partition function can be computed perturbatively in $1/k$ by expanding the action to quadratic order around the saddle points. 

One expands the gauge potential around a flat connection $A^{(\ell)}$,  
\bea
A_m &=& A^{(\ell)}_m + a_m
\eea
Since we are not interested in gauge transforming the flat connection to zero (if we do that, then we change the boundary conditions of the fields), it is natural to impose the following gauge transformation rules for these new fields,
\bea
{A^{(\ell)}}^g_m &=& g^{-1} A^{(\ell)}_m g\cr
a^g_m &=&  i g^{-1} \nabla_m g + g^{-1} a_m g 
\eea
meaning that we can only rotate the flat connection (in particular we can diagonalize it), but not gauge transform it to zero. On the other hand, the fluctuation field is now a gauge potential that we need to gauge fix. We define a derivative
\bea
D^{(\ell)}_m := \nabla_m - i[A^{(\ell)},\cdot]
\eea
and consider the following nilpotent BRST variations
\bea
\delta a_m &=& D_m^{(\ell)} c\cr
\delta A^{(\ell)}_m &=& 0\cr
\delta c &=& 0\cr
\delta B &=& 0\cr
\delta \b{c} &=& i B
\eea
We add the gauge fixing term
\bea
S_{gauge}(a) &=& -i\delta \int d^3 x \sqrt{g} \tr\(\b{c} D^{(\ell)m} a_m\) = \int d^3 x \sqrt{g} \tr\(B D^{(\ell) m} a_m + i \b{c} D^{(\ell)m} D^{(\ell)}_m c\)
\eea
which we will write as
\bea
S_{gauge}(a) &=& (B,D^{(\ell) \dag} a) + i (\b{c},\triangle^{(\ell)}_0 c)
\eea
We now see that we could also have used the original BRST variations and the full covariant derivative $D_m$. Then we would get the same gauge fixing action $S_{gauge}$ with higher order correction terms, which would play no role for the 1-loop computation.

By multiplying $S_{gauge}$ by an overall constant $\frac{k}{2\pi}$, the full BRST gauge fixed action becomes of the form
\bea
S(a,B,c,\b c) &=& S(A^{(\ell)}) + \frac{k}{4\pi} (a,*Da) + \frac{k}{2\pi} \[(B,D^{\dag}a) + i (\b{c},\triangle_0 c)\]
\eea
We can write part of this action as
\bea
\frac{k}{4\pi} \(\begin{pmatrix} a & B \end{pmatrix}, 
\begin{pmatrix} *D & D\\ D^{\dag} & 0 \end{pmatrix}
\begin{pmatrix} a \\ B \end{pmatrix}\)
\eea
The matrix operator that enters in this expression has the square 
\ben
\begin{pmatrix} *D & D\\ D^{\dag} & 0 \end{pmatrix}
\begin{pmatrix} *D & D\\ D^{\dag} & 0 \end{pmatrix} &=& \begin{pmatrix} \triangle_1 & 0\\ 0 & \triangle_0 \end{pmatrix}\label{square}
\een
One may also notice that the operator we just squared, is nothing but $L_-$, which is defined from $L = *D + D*$ by restriction to odd forms. If $f_1$ and $f_3$ denote a one-form and a three-form, with coefficients $a_1$ and $a_3$, then we find that 
\bea
L_- (a_1 f_1 + a_3 f_3) &=& \(a_1 * Df_1 + a_3 D*f_3\) + a_1 D*f_1
\eea
If we then write $f_0 = *f_3$, then we find that 
\bea
L_- \begin{pmatrix} f_1 \\ f_0 \end{pmatrix} &=& \begin{pmatrix} *D & D\\ D^{\dag} & 0 \end{pmatrix}\begin{pmatrix} f_1 \\ f_0 \end{pmatrix}
\eea
The contribution from the flat connection $A^{(\ell)}$ to the partition function becomes
\bea
\exp S(A^{(\ell)}) \frac{\det\triangle_0}{\sqrt{\det(L_-)}}
\eea
Moreover, 
\bea
\frac{1}{\sqrt{\det L_-}} &=& \frac{1}{\sqrt{|\det L_-|}} \exp \frac{i\pi}{2} \eta(A^{(\ell)})
\eea
where, from the APS index theorem,
\bea
\frac{1}{2} \eta(A^{(\ell)}) &=& \frac{1}{2} \eta(0) + \frac{c_2}{2\pi} S(A^{(\ell)})
\eea
Thus this phase factor can be absorbed by shifting 
\bea
k \rightarrow K:=k+c_2/2
\eea

Let us return to the absolute value of the partition function. From (\ref{square}) together with the ghost contribution, we get
\bea
\frac{\det\triangle_0}{\sqrt{|\det(L_-)|}} = \(\det \triangle_0\)^{\frac{3}{4}} \(\det \triangle_1\)^{-\frac{1}{4}} 
\eea
If we take away the zero modes, this is the oscillator mode contribution to the square root of the RS torsion.

\subsection{The dependence on the Chern-Simons level}
To derive the $K$-dependence, all we need to do, is to extract the $K$-dependence from the kinetic term inside the Chern-Simons term. The path integral gives the factor
\bea
\frac{1}{\det \(K *d\)^{\frac{1}{2}}} &=& \frac{1}{K^{\frac{1}{2}\zeta_{*d}(0)}} \frac{1}{\det\(*d\)^{\frac{1}{2}}}\cr
\zeta_{*d}(0) &=& b_0 - b_1
\eea
If we assume that $b_1 = 0$, there will be no bosonic zero modes of the operator $d$ and the zero mode problem can be avoided. And then this gives the correct $K$-dependence. More specifically, $b_q$ is the dimension of $H_q(M_3)$ times the dimension of the unbroken gauge group in the background of the flat connection $A^{(\ell)}$.

Let us finally review the computation of the perturbative partition function for $G = SU(2)$ gauge group on lens space $S^3/\mb{Z}_p$. There are flat connections $A^{(\ell)}$
for $\ell = 0,1,2,...,p-1$. We shall divide by the isotropy group of unbroken gauge symmetries when we turn on the flat connection. When $\ell=0$  the isotropy group is $H_{A^{(0)}}=SU(2)$ as no gauge symmetry is broken. When $\ell>0$ the isotropy group is $H_{A^{(\ell)}}=U(1)$. 

Since the classical Chern-Simons action is normalized as
\bea
\frac{ik}{2\pi} \frac{1}{2}(A,*dA)
\eea
which is off the canonical normalization by the factor of $\frac{iK}{2\pi}$, the perturbative computation of the path integral will give the result (assuming that $p$ is odd)
\bea
Z &=& \frac{\(\frac{iK}{2\pi}\)^{-\frac{1}{2} b_0(A^{(0)})}}{\Vol\(H_{A^{(0)}}\)}  \tau(A^{(0)})^{\frac{1}{2}} + \sum_{\ell=1}^{\frac{p-1}{2}} \frac{\(\frac{iK}{2\pi}\)^{-\frac{1}{2} b_0(A^{(\ell)})}}{\Vol\(H_{A^{(\ell)}}\)}  e^{2\pi i K \ell^2/p} \tau(A^{(\ell)})^{\frac{1}{2}} 
\eea
We restrict the sum to run over $\ell = 1,...,(p-1)/2$ following Eq (2.18) in \cite{Freed:1991wd}, Eq (4.17) in \cite{Adams:1995xj}, and \cite{Rozansky:1993zx}. Here the RS torsions are given by 
\bea
\tau(A^{(0)}) &=& \(\frac{1}{p}\)^3
\eea
the power 3 because there are three generators of $SU(2)$, and 
\bea
\tau(A^{(\ell)}) &=& \frac{1}{p} \(2\sin\frac{2\pi\ell}{p}\)^4
\eea
The volume of $SU(2)=S^3$ with unit radius $r$ is $\Vol(SU(2)) = 2\pi^2 r^3$ and the length of the equator is $\Vol(U(1)) = 2\pi r$. 

Using $b_0(A^{(0)}) = 3$ and $b_0(A^{(\l)})=1$, and by comparing with the known exact result to be presented in below, we get
\bea
\Vol(SU(2)) &=& 2 \sqrt{\pi}\cr
\Vol(U(1)) &=& 2 \sqrt{\pi}
\eea
This is consistent with taking the radius of $SU(2)$ as
\bea
r &=& \frac{1}{\sqrt{\pi}}
\eea

\subsection{The exact result}
For $SU(2)$ gauge group, the exact result for the partition function on $L(p;1)$ is given by
\bea
Z(\epsilon) &=& - e^{\frac{3\pi i}{4}-\frac{i(3-p)\epsilon}{4}} \sum_{\l=0}^{p-1} \int_{C^{(\l)}} \frac{dz}{2\pi i} \sinh^2 \(\frac{z}{2}\) e^{\frac{i p z^2}{4\epsilon} - \frac{2\pi \l}{\epsilon} z} 
\eea
where $C^{(\l)}$ is the contour 
\bea
z &=& e^{\frac{i\pi}{4}} x - \frac{4\pi i\l}{p}
\eea
for $\l = 0,...,p-1$. (This result can be extracted from Eq (5.38) in \cite{Beasley:2005vf} by taking $P=1$, $N=0$, $d=p$ and $\theta_0 = 3-p$ in the expression there.) Here
\bea
\epsilon &=& \frac{2\pi}{k+2}
\eea
where $k+2$ is the shifted Chern-Simons level. 

For $p =1$ the formula reproduces the famous result \cite{Witten:1988hf}
\bea
Z(S^3) &=& \sqrt{\frac{2}{k+2}} \sin \(\frac{\pi}{k+2}\)
\eea
for the partition function on $S^3$. 

For generic $p$, the integrals can also be computed exactly with the following result
\bea
Z(\epsilon) &=& \frac{1}{2 i} \sqrt{\frac{\epsilon}{\pi p}} e^{\frac{\pi i}{4} (p-3)} \sum_{\l=0}^{p-1} e^{\frac{4\pi^2 i}{\epsilon} \frac{\l^2}{p}} \(e^{\frac{i \epsilon}{p}} \cos \frac{4\pi \l}{p} - 1\)
\eea

To also see the shift from $k$ to $k+2$ we would need to compute the eta invariant.

Because the $\l=0$ term has different leading term $K$ asymptotics from the terms with $\l>0$, we separate the sum into these two pieces and pick up only the leading term from each piece 
\bea
Z_0(\epsilon) &=& \frac{1}{2 i} \sqrt{\frac{\epsilon}{\pi p}} e^{\frac{\pi i}{4} (p-3)} \(e^{\frac{i \epsilon}{p}} - 1\)\cr
&=& e^{\frac{\pi i}{4}p} \sqrt{2} \pi \(\frac{1}{i K p}\)^{3/2} 
\eea
and 
\bea
Z_{rest}(\epsilon) &=& \frac{1}{2 i} \sqrt{\frac{\epsilon}{\pi p}} e^{\frac{\pi i}{4} (p-3)} \sum_{\l=1}^{p-1} e^{\frac{4\pi^2 i}{\epsilon} \frac{\l^2}{p}} \(\cos \frac{4\pi \l}{p} - 1\)\cr
&=& e^{\frac{\pi i}{4}p} 2 \sqrt{\frac{1}{2 i K p}} \sum_{\l=1}^{p-1} e^{\frac{2\pi i K \l^2}{p}} \(\sin \frac{2\pi \l}{p}\)^2\cr
\eea
There will be an order $K^{-3/2}$ contribution to $Z_{rest}$ as well, but we can ignore that since each $\l$-sector can be studied on its own. By a complex conjugation $i\rightarrow -i$, we have now obtained Eq (2.37) in \cite{Rozansky:1993zx}.

The result in \cite{Rozansky:1993zx} was presented for odd $p$. In that case, the sum can be replaced by twice of half of the sum as
\bea
\sum_{\l=1}^{p-1} e^{\frac{2\pi i K \l^2}{p}} \(\sin \frac{2\pi \l}{p}\)^2 = 2 \sum_{\l=1}^{\frac{p-1}{2}} e^{\frac{2\pi i K \l^2}{p}}\(\sin \frac{2\pi \l}{p}\)^2
\eea
On the other hand, if $p$ is even, then the RS torsion is vanishing for $\ell=p/2$ and we can write
\bea
\sum_{\l=1}^{p-1} e^{\frac{2\pi i K \l^2}{p}} \(\sin \frac{2\pi \l}{p}\)^2 = 2 \sum_{0<\l<p/2} e^{\frac{2\pi i K \l^2}{p}}\(\sin \frac{2\pi \l}{p}\)^2
\eea
in agreement with \cite{Freed:1991wd}.

\section{Dimensional reduction of selfdual forms on a circle}\label{6dto5d}
We consider a nonselfdual $2k$-form potential on Euclidean $S^1\times M_{4k+1}$. The ghost hierarchi grows linearly, which gives the partition functions as
\bea
Z_{4k+2} &=& \frac{\prod_{\l=1}^{k}\(\det{}'\triangle_{2k-2\l+1}\)^{\frac{2\l}{2}}}{\prod_{\l=0}^k\(\det{}'\triangle_{2k-2\l}\)^{\frac{2\l+1}{2}}}
\eea
The partition function of a $(2k-1)$-potential on $M_{4k+1}$ is likewise given by
\bea
Z_{4k+1} &=& \frac{\prod_{\l=1}^k \(\det{}'\triangle_{2k-2\l}\)^{\frac{2\l}{2}}}{\prod_{\l=1}^k \(\det{}'\triangle_{2k-2\l+1}\)^{\frac{2\l-1}{2}}}
\eea
We then dimensionally reduce along $S^1$ by replacing $\det{}'\triangle_p$ by $\det{}'\triangle_p \det{}'\triangle_{p-1}$ where the latter represent Laplacians on $M_{4k+1}$. This gives the dimensionally reduced partition function as
\bea
Z_{4k+2}(\partial_{\tau}=0) &=& \frac{\prod_{\l=1}^{k}\(\det{}'\triangle_{2k-2\l+1}\)^{\frac{1}{2}}}{\prod_{\l=0}^k\(\det{}'\triangle_{2k-2\l}\)^{\frac{1}{2}}}
\eea
We then compute the ratio
\bea
\frac{Z_{4k+2}(\partial_{\tau}=0)}{Z_{4k+1}^2} &=& \frac{\prod_{\l=1}^{k}\(\det{}'\triangle_{2k-2\l+1}\)^{2\l-\frac{1}{2}}}{\prod_{\l=0}^k\(\det{}'\triangle_{2k-2\l}\)^{2\l+\frac{1}{2}}}
\eea
Using Poincare duality, we get the Ray-Singer torsion on a $(4k+1)$-dimensional manifold $M_{4k+1}$ as
\bea
\tau_{osc}(M_{4k+1}) &=& \prod_{p=0}^{2k} \(\det{}'\triangle_p\)^{(-1)^p \frac{4k+1-2p}{2}}\cr
&=& \frac{\prod_{\l=0}^k \(\det{}'\triangle_{2k-2\l}\)^{2\l+\frac{1}{2}}}{\prod_{\l=1}^{k} \(\det{}'\triangle_{2k-2\l+1}\)^{2\l-\frac{1}{2}}}
\eea
and we see that 
\bea
\frac{Z_{4k+2}(\partial_{\tau}=0)}{Z_{4k+1}^2} &=& \frac{1}{\tau_{osc}(M_{4k+1})}
\eea
holds for any $k=0,1,2,3,...$. For the case $k=0$ which corresponds to a zero-form in 2d, the relation still holds if we assume that the 1d oscillator partition function is equal to one. 

In $4k-1$ dimensions, the analytic torsion is
\bea
\tau_{osc} &=& \frac{\prod_{\l=1}^k \(\det{}'\triangle_{2k-2\l}\)^{{2\l-\frac{1}{2}}}}{\prod_{\l=0}^{k-1} \(\det{}'\triangle_{2k-2\l-1}\)^{{2\l+\frac{1}{2}}}}
\eea
We have
\bea
Z_{4k} &=& \frac{\prod_{\l=1}^k \(\det{}'\triangle_{2k-2\l}\)^{\frac{2\l}{2}}}{\prod_{\l=0}^{k-1} \(\det{}'\triangle_{2k-2\l-1}\)^{\frac{2\l+1}{2}}}
\eea
and
\bea
Z_{4k-1} &=& \frac{\prod_{\l=0}^{k-1} \(\det{}'\triangle_{2k-2\l-1}\)^{\frac{2\l}{2}}}{\prod_{\l=0}^{k-1} \(\det{}'\triangle_{2k-(2\l+2)}\)^\frac{2\l+1}{2}}
\eea
Then 
\bea
Z_{4k}(\partial_{\tau}=0) &=& \frac{\prod_{\l=1}^{k} \(\det{}'\triangle_{2k-2\l}\)^{\frac{1}{2}}}{\prod_{\l=0}^{k-1}\(\det{}'\triangle_{2k-2\l=1}\)^{\frac{1}{2}}}
\eea
Then the relation instead becomes
\bea
\frac{Z_{4k}(\partial_{\tau}=0)}{Z_{4k-1}^2} &=& \tau_{osc}(M_{4k})
\eea



\begin{thebibliography}{100}




\bibitem{Witten:1988hf}
  E.~Witten,
  ``Quantum Field Theory and the Jones Polynomial,''
  Commun.\ Math.\ Phys.\  {\bf 121} (1989) 351.



\bibitem{Beasley:2005vf}
  C.~Beasley and E.~Witten,
  ``Non-Abelian localization for Chern-Simons theory,''
  J.\ Diff.\ Geom.\  {\bf 70} (2005) no.2,  183
  [hep-th/0503126].


\bibitem{Blau:2006gh}
  M.~Blau and G.~Thompson,
  ``Chern-Simons theory on S1-bundles: Abelianisation and q-deformed Yang-Mills theory,''
  JHEP {\bf 0605} (2006) 003
  [hep-th/0601068].


\bibitem{Kallen:2011ny}
  J.~Kallen,
  ``Cohomological localization of Chern-Simons theory,''
  JHEP {\bf 1108} (2011) 008
  [arXiv:1104.5353 [hep-th]].


\bibitem{Witten:1992xu}
  E.~Witten,
  ``Two-dimensional gauge theories revisited,''
  J.\ Geom.\ Phys.\  {\bf 9} (1992) 303
  [hep-th/9204083]. 

\bibitem{Pestun:2007rz} 
  V.~Pestun,
  ``Localization of gauge theory on a four-sphere and supersymmetric Wilson loops,''
  Commun.\ Math.\ Phys.\  {\bf 313}, 71 (2012)
  [arXiv:0712.2824 [hep-th]].


\bibitem{Freed:1991wd} 
  D.~S.~Freed and R.~E.~Gompf,
  ``Computer calculation of Witten's three manifold invariant,''
  Commun.\ Math.\ Phys.\  {\bf 141}, 79 (1991).


\bibitem{Rozansky:1993zx}
  L.~Rozansky,
  ``A Large k asymptotics of Witten's invariant of Seifert manifolds,''
  Commun.\ Math.\ Phys.\  {\bf 171} (1995) 279
  [hep-th/9303099].

\bibitem{Adams:1995xj} 
  D.~H.~Adams and S.~Sen,
  ``Partition function of a quadratic functional and semiclassical approximation for Witten's three manifold invariant,''
  hep-th/9503095.

\bibitem{Adams:1995np} 
  D.~H.~Adams and S.~Sen,
  ``Phase and scaling properties of determinants arising in topological field theories,''
  Phys.\ Lett.\ B {\bf 353}, 495 (1995)
  [hep-th/9506079].

\bibitem{Adams:1996hi} 
  D.~H.~Adams,
  ``A Note on the Faddeev-Popov determinant and Chern-Simons perturbation theory,''
  Lett.\ Math.\ Phys.\  {\bf 42}, 205 (1997)
  [hep-th/9704159].

\bibitem{Adams:1997zc}
  D.~H.~Adams,
  ``The semiclassical approximation for the Chern-Simons partition function,''
  Phys.\ Lett.\ B {\bf 417} (1998) 53
  [hep-th/9709147].


\bibitem{Blau:1989bq}
  M.~Blau and G.~Thompson,
  ``Topological Gauge Theories of Antisymmetric Tensor Fields,''
  Annals Phys.\  {\bf 205} (1991) 130.
  


\bibitem{Bak:2015hba}
  D.~Bak and A.~Gustavsson,
  ``The geometric Langlands twist in five and six dimensions,''
  JHEP {\bf 1507} (2015) 013
  [arXiv:1504.00099 [hep-th]].

\bibitem{Park}
J-H. Park, N. Nekrasov, private notes


\bibitem{Ray}
D.B Ray,
``Reidemeister torsion and the laplacian on lens spaces,''
Advances in Mathematics 4, 109-126 (1970)

 
\bibitem{Nash:1992sf}
  C.~Nash and D.~J.~O'Connor,
  ``Determinants of Laplacians, the Ray-Singer torsion on lens spaces and the Riemann zeta function,''
  J.\ Math.\ Phys.\  {\bf 36} (1995) 1462
   Erratum: [J.\ Math.\ Phys.\  {\bf 36} (1995) 4549]
  [hep-th/9212022].
 
\bibitem{Friedmann:2002ty}
  T.~Friedmann and E.~Witten,
  ``Unification scale, proton decay, and manifolds of G(2) holonomy,''
  Adv.\ Theor.\ Math.\ Phys.\  {\bf 7} (2003) no.4,  577
  [hep-th/0211269].


\bibitem{Bunke}
U.~Bunke,~``Lectures on analytic torsion,''~http://www.uni-regensburg.de/Fakultaeten/nat\_Fak\_I/Bunke/sixtorsion.pdf 

\bibitem{Mnev}
P.~Mnev,~ ``Lecture notes on torsions,''~eprint arXiv:1406.3705.


\bibitem{Imamura:2013qxa} 
  Y.~Imamura, H.~Matsuno and D.~Yokoyama,
  ``Factorization of the $S^3/\mathbb{Z}_n$ partition function,''
  Phys.\ Rev.\ D {\bf 89}, no. 8, 085003 (2014)
  [arXiv:1311.2371 [hep-th]].

\bibitem{Kirk}
P. A. Kirk,  
 and E. P. Klassen, 
 "Chern-Simons invariants of 3-manifolds and representation spaces of knot groups.." Mathematische Annalen 287.2 (1990): 343-368. http://eudml.org/doc/164690.



\bibitem{Witten:1991we}
  E.~Witten,
  ``On quantum gauge theories in two-dimensions,''
  Commun.\ Math.\ Phys.\  {\bf 141} (1991) 153.

\bibitem{Blau:1993tv}
  M.~Blau and G.~Thompson,
  ``Derivation of the Verlinde formula from Chern-Simons theory and the G/G model,''
  Nucl.\ Phys.\ B {\bf 408} (1993) 345
  doi:10.1016/0550-3213(93)90538-Z
  [hep-th/9305010].
  
\bibitem{Marino:2011nm} 
  M.~Marino,
  ``Lectures on localization and matrix models in supersymmetric Chern-Simons-matter theories,''
  J.\ Phys.\ A {\bf 44}, 463001 (2011)
  [arXiv:1104.0783 [hep-th]].


\bibitem{Kallen:2012cs}
  J.~K{a}llen and M.~Zabzine,
  ``Twisted supersymmetric 5D Yang-Mills theory and contact geometry,''
  JHEP {\bf 1205} (2012) 125
  [arXiv:1202.1956 [hep-th]].
  




\bibitem{Guadagnini:2014mja}
  E.~Guadagnini and F.~Thuillier,
  ``Path-integral invariants in abelian Chern-Simons theory,''
  Nucl.\ Phys.\ B {\bf 882} (2014) 450
  [arXiv:1402.3140 [hep-th]].


\bibitem{Siegel:1980jj} 
  W.~Siegel,
  ``Hidden Ghosts,''
  Phys.\ Lett.\  {\bf 93B}, 170 (1980).


\bibitem{Kimura:1980aw} 
  T.~Kimura,
  ``Quantum Theory of Antisymmetric Higher Rank Tensor Gauge Field in Higher Dimensional Space-time,''
  Prog.\ Theor.\ Phys.\  {\bf 65}, 338 (1981).
  



\bibitem{Douglas:2010iu}
  M.~R.~Douglas,
  ``On D=5 super Yang-Mills theory and (2,0) theory,''
  JHEP {\bf 1102} (2011) 011
  [arXiv:1012.2880 [hep-th]].
  
\bibitem{Lambert:2010iw}
  N.~Lambert, C.~Papageorgakis and M.~Schmidt-Sommerfeld,
  ``M5-Branes, D4-Branes and Quantum 5D super-Yang-Mills,''
  JHEP {\bf 1101} (2011) 083
  [arXiv:1012.2882 [hep-th]].



\bibitem{Bak:2016vpi} 
  D.~Bak and A.~Gustavsson,
  ``Witten indices of abelian M5 brane on $ \mathbb{R}\times {S}^5 $,''
  JHEP {\bf 1611}, 177 (2016)
  [arXiv:1610.06255 [hep-th]].




\bibitem{b}
Theorem 5.2 in `The Laplacian on a Riemannian Manifold: An Introduction to Analysis on Manifolds', Steven Rosenberg, Cambridge University Press, 1997














    
\end{thebibliography}
\end{document}